\documentclass[superscriptaddress,reprint,amsmath,amssymb,aps,epsf,prb,twocolumn]{revtex4-2}
\usepackage[utf8]{inputenc}
\usepackage{graphicx}
\usepackage{tikz}
\usepackage{dcolumn}
\usepackage{bm}
\usepackage{amsmath}
\usepackage{amsfonts}
\usepackage{amssymb}
\usepackage{hyperref}
\usepackage{pbox}
\usepackage{pinlabel}
\usepackage{ragged2e}
\usepackage[justification=RaggedRight]{caption}
\usepackage[singlelinecheck=off]{subcaption}
\usepackage{ulem}

\newcommand{\kvec}{\boldsymbol{k}}
\newcommand{\psivec}{\boldsymbol{\psi}}

\begin{document}

	\title{Interacting nodal semimetals with non-linear bands}

 \author{Arianna Poli}
	\affiliation{Dipartimento di Scienze Fisiche e Chimiche, Università dell’Aquila, Coppito-L’Aquila, Italy}	
	\author{Niklas Wagner}
	\affiliation{Institut f\"ur Theoretische Physik und Astrophysik, Universit\"at W\"urzburg, 97074 W\"urzburg, Germany}
 	\author{Max Fischer}
	\affiliation{Institut f\"ur Theoretische Physik und Astrophysik, Universit\"at W\"urzburg, 97074 W\"urzburg, Germany}
	\author{Alessandro Toschi}
	\affiliation{Institute of Solid State Physics, TU Wien, 1040 Vienna, Austria}
	\author{Giorgio Sangiovanni}
	\affiliation{Institut f\"ur Theoretische Physik und Astrophysik and W\"urzburg-Dresden Cluster of Excellence ct.qmat, Universit\"at W\"urzburg, 97074 W\"urzburg, Germany}
	\author{Sergio Ciuchi}%
	\affiliation{Dipartimento di Scienze Fisiche e Chimiche, Università dell’Aquila, Coppito-L’Aquila, Italy}
\affiliation{Istituto dei Sistemi Complessi, CNR, 00185 Roma, Italy}

 \begin{abstract}

We investigate the quasi-particle and transport properties of a model describing interacting Dirac and Weyl semimetals in the presence of local Hubbard repulsion $U$, where we explicitly include a deviation from the linearity of the energy-momentum dispersion through an intermediate-energy scale $\Lambda$. Our focus lies on the correlated phase of the semimetal. 

At the nodal point, the renormalization of spectral weight at a fixed temperature $T$ exhibits a weak dependence on $\Lambda$ but is sensitive to the proximity to the Mott transition. Conversely, the scattering rate of quasi-particles and the resistivity display high-temperature exponents that crucially rely on $\Lambda$, leading to a crossover towards a conventional Fermi-liquid behaviour at finite T. Finally, by employing the Nernst-Einstein relation for conductivity, we identify a corresponding density crossover as a function of the chemical potential.

\end{abstract}

\maketitle
		
\noindent
\section{Introduction}
\label{sec:intro}

Nodal semimetals are characterized by a linear crossing of bands, protected either by topology or by lattice symmetries. Contrary to conventional metals possessing a sizeable density of the states (DOS) at the Fermi level, the DOS of nodal semimetals in three (two) spatial dimensions, vanishes quadratically (linearly). The consequently small carriers' density at the nodal point is the origin of non-trivial transport properties. 
The possibility of having excitations with arbitrarily small energy gives indeed rise to non-trivial resistivity temperature exponents \citep{hosur_charge_2012} \citep{burkov_topological_2011}.

\noindent
Further, the shrinking of the Fermi surface down to a single point reduces the phase space for fermion-fermion scattering and guarantees robust protection against electronic correlations for short-range interactions \citep{coleman_2015,nozieres1999theory}. In the weak-to-intermediate regime, the lifetime of quasi-particles is indeed much longer than in conventional Fermi liquids. This shielding effect breaks down only at strong coupling, where the self-energy can diverge and a Mott insulating state sets in.

\begin{figure}[b]
\centering
\begin{minipage}[b]{0.49\textwidth}
\includegraphics[scale=0.12]{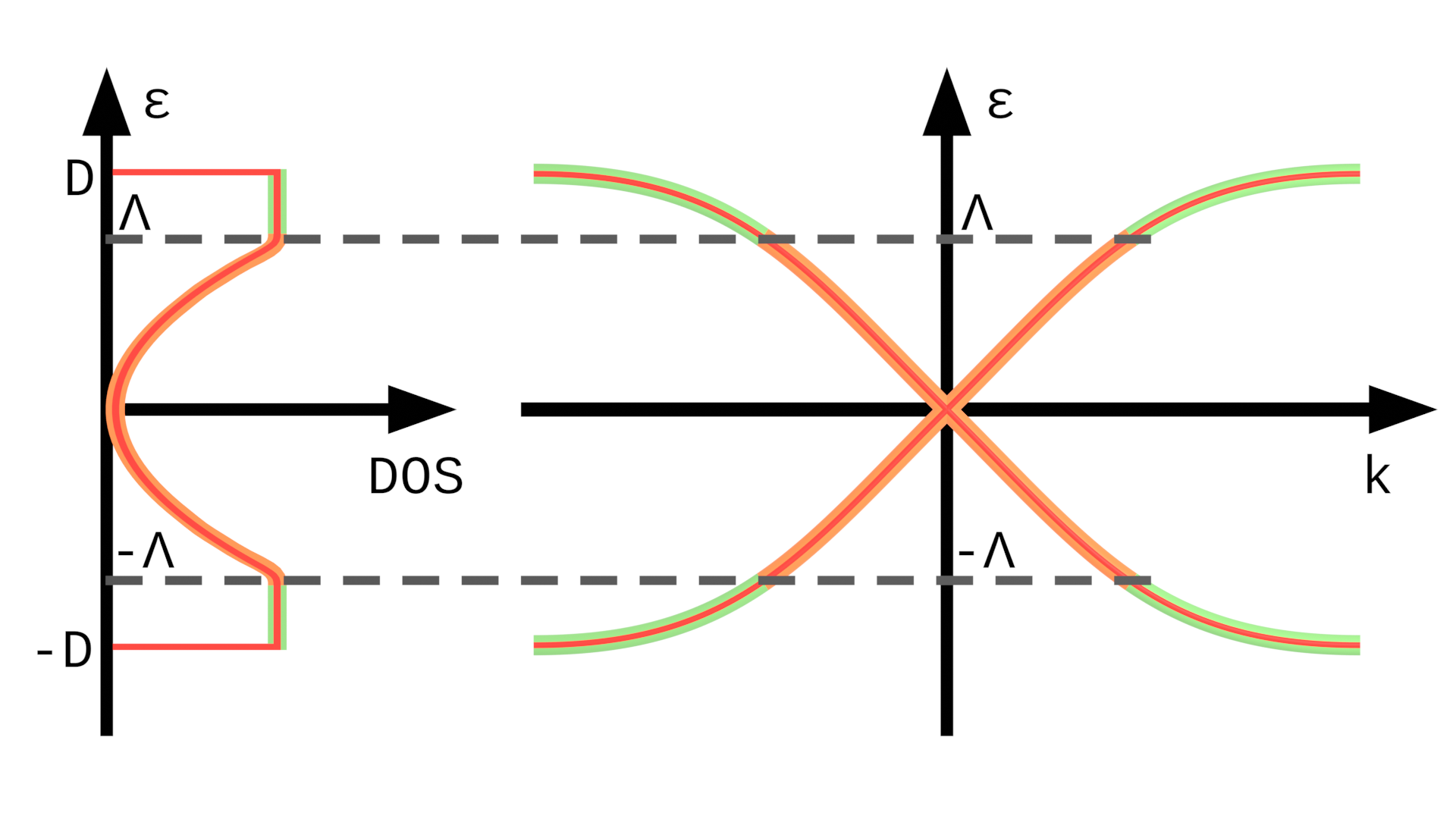}
\caption{The free particle energy dispersion (on the right) and the correspondent DOS in 3 dimensions (on the left) are shown. The DOS is taken quadratic at low energy $|\epsilon| < \Lambda$ and constant at high energy  $\Lambda<|\epsilon| <D$. The cutoff $\Lambda$ sets the scale where the deviation from the linearity of the bands becomes relevant.}
\label{fig:DOSdiracsquare}
\end{minipage}
\hfill
\end{figure}

From a general point of view
the electronic properties of realistic semimetals are significantly affected by the precise spin/orbital structure of the band dispersion in the whole vicinity of the nodal point. In particular, the position of the Fermi energy with respect to the symmetry point significantly impacts the low-temperature behaviours. 
Further, corrections beyond the considered ``Lorentz-invariant'' linear dispersion translate into corresponding deviations in the DOS at a scale -- defined henceforth as $\Lambda$ -- at which the non-linearity of the dispersion starts to appear (see the sketch in Fig.~\ref{fig:DOSdiracsquare}). 

The goal of the present work is to study the competition between two effects: the protection against electron-electron interaction originating from the fermiology of the semimetallic band structure on the one hand and the tendency of the system to flow to strong coupling when the strength of the local Hubbard interaction $U$ reaches the electronic half-bandwidth $D$. The interplay between these two counteracting effects is enriched by the presence of the intermediate scale of non-linear dispersion $\Lambda$ and by finite doping when the chemical potential no longer sits at the Dirac cone. Both are non-universal factors which are present in materials and quantitatively influence their behaviour and are expected to change, for instance, the transport exponents as well as the position of the crossover scales in the phase diagram of interacting semimetals. In particular, we will focus on quasiparticle properties and transport as a function of the temperature and of the Hubbard repulsion for different $\Lambda/D$ ratios. 

We need to go beyond perturbative approaches in the electron-electron interaction. 
For this reason in order to address weakly and strongly interacting semimetals on equal footing, we apply dynamical mean field theory (DMFT) \citep{georges_dynamical_1996}. DMFT maps the many-electron lattice model onto a self-consistent quantum impurity problem which, in this case, consists of two orbitals and two spin species.
In three spatial dimensions, DMFT can be viewed as a tool for getting an approximate momentum-independent self-energy. The latter is obtained by solving the DMFT equations written in terms of the local Green's function only. Further, the momentum-dependent Green's function can be calculated with the Hamiltonian describing the non-interacting lattice model under consideration. For this, we will use here a corresponding DOS sketched in Fig.~\ref{fig:DOSdiracsquare} which is specific for the 3D case and where the half-bandwidth $D$ is our energy unit.

As we describe in Sec.~\ref{sec:model}, we solve the DMFT equations within iterated perturbation theory (IPT), which keeps enough accuracy on the real frequency axis \citep{georges_hubbard_1992,kajueter_new_1996,kajueter_doped_1996} and hence allows us to reliably extract transport quantities as done in Ref.~\citep{wagner}.
Apart from the specific solver, a general disclaimer on single-site DMFT is in order, particularly when focusing on transport properties: nodal points are well protected against local interactions. Single-site DMFT does not include a momentum-dependent self-energy therefore the long wavelength effects of interaction are absent. In the case of a vanishing DOS, these effects are relevant while approaching the nodal point and the value of the temperature exponents can be modified with respect to the single site DMFT \citep{MillisAnqiPRB2022}.

\section{The model}
\label{sec:model}
We consider the following Hamiltonian for electrons ($\sigma=\uparrow, \downarrow$) in two orbitals ($\alpha=1,2$) interacting through on-site Hubbard repulsion 
\begin{equation}
    \hat{H}=\sum_{\kvec} \psivec^\dagger_{\kvec} \boldsymbol{\hat{H}_0}^{(4\times4)} (\kvec) \psivec_{\kvec}+ U\sum_{i,\alpha}  (\boldsymbol{n}_{i,\alpha,\uparrow}-\frac{1}{2})(\boldsymbol{n}_{i,\alpha,\downarrow}-\frac{1}{2})
    \label{eq:Hamiltonian}
\end{equation}
where $\psivec^{(\dagger)}_{\kvec}\equiv(c^{(\dagger)}_{\kvec,1,\uparrow},c^{(\dagger)}_{\kvec,2,\uparrow},c^{(\dagger)}_{\kvec,1,\downarrow},c^{(\dagger)}_{\kvec,2,\downarrow})$ is a four-component spinor of annihilation (creation) operators for the electrons in $\kvec-$space. The density of carriers $\boldsymbol{n}_{i,\alpha,\sigma}=c^{\dagger}_{i,\alpha,\sigma}c_{i,\alpha,\sigma}$.

The matrix $\boldsymbol{H_0} (\kvec)$ defines the non-interacting part of the Hamiltonian that can have the general form

\begin{equation}
    \boldsymbol{\hat{H}}_0^{(4\times4)}(\kvec)=\vec{\sigma}_z\otimes\vec{b}(\kvec)\cdot\vec{\tau}
    \label{eq:Hamiltonian_block}
\end{equation}
where the energy dispersion function is
\begin{equation}
    \vec{b}(\kvec)=(b_x(\kvec),b_y(\kvec),M(\kvec))
    \label{eq:b_vector}
\end{equation}

Here the Pauli matrices $\vec{\sigma}$ act on spin and $\vec{\tau}$ on orbital space. The three components of $\Vec{b}(\kvec)$ are such that around the Dirac node (here located at $\kvec = 0$), they acquire a linear $\kvec$-dependence such that for an isotropic 
Dirac node, we have $\boldsymbol{\hat{H}_0}^{(4\times4)}(\kvec)\simeq\vec{\sigma}_z\otimes v_F\kvec \cdot\vec{\tau}$ with $v_F$ being the Fermi velocity.

Since we do not consider spontaneous time-reversal symmetry-breaking and our microscopic Hamiltonian does not contain explicit time-reversal-breaking terms, all calculations shown in the following will be representative of both Dirac and Weyl fermions.
The interplay between the phase-space suppression of electron-electron (el-el) scattering and local Mottness is mainly determined by the form of the band structure and on the type of interaction. For this reason, this basic competition of physical effects would equally characterize actual Weyl fermions in the presence of time-reversal breaking terms in the single-particle Hamiltonian. The decisive point is the parabolic density of the states of the three-dimensional lattice with linear dispersion.

The second term in Eq. (\ref{eq:Hamiltonian}) is a purely local Hubbard repulsion. We use DMFT to treat this last term. 
For a general Hamiltonian of the kind of Eq. (\ref{eq:Hamiltonian_block}) the standard DMFT procedure  gives a local self-energy which, in view of the diagonal form of interaction in the orbital space, is also diagonal.  In this work, we analyze the semi-metallic correlated and Mott phases in the absence of long-range order. In real-life materials, such states are reachable upon heating above the transition temperatures, in particular when these are suppressed by lattice geometrical effects \citep{https://doi.org/10.1002/pssb.201248476,RevModPhys.70.1039}. Paramagnetic DMFT is hence an ideal way of studying the effect of strong frustration.  

Together with the action of local interaction we take into account deviations from quadratic low-energy behaviour of the total DOS which are included in the specific form of the vector $\Vec{b}(\kvec)$. 
In particular, we introduce an intermediate energy cut-off $\Lambda$  separating different zones in energy space as is depicted in Fig.~\ref{fig:DOSdiracsquare} (see Appendix \ref{app:DMFT} for details).

In a low energy sector ($|\epsilon|< \Lambda$), we will consider a linear approximation for the band structure in Eq. (\ref{eq:Hamiltonian_block}). 
For ($|\epsilon|>\Lambda$) instead we will neglect $b_x(\kvec),b_y(\kvec)$ terms in Eq. (\ref{eq:Hamiltonian_block}) to maintain only the mass term. In addition assuming that the orbital-diagonal term $M(\kvec)$ is such that $\sum_{\kvec} M(\kvec)=0$  we arrive to an expression of the local Green's function which is a functional of the total DOS (see Appendix \ref{app:DMFT}).

This choice takes into account not only finite-bandwidth effects ($|\omega|<D$) essential to deal with the metal-insulator transition as in Ref.~\citep{wagner}, but also deviations from linearity occurring above the intermediate energy cutoff $\Lambda$.
A flat DOS is considered to mimic a more realistic one as that obtained from tight-binding models (see e.g. Ref.~\citep{KargarianPNAS2016}) at intermediate frequencies $\Lambda<|\omega|<D$.
This form of the DOS will allow for analytical estimates of various quantities of interest.

\section{The semimetal Insulator Transition at nodal point}
\label{sec:sMIT}

We describe here the approach to the semimetal-to-insulator transition (sMIT) that occurs when the chemical potential is fixed at the nodal point upon increasing the interaction's strength $U$. The overall behaviour is qualitatively similar to the prototypical Mott transition in DMFT for a conventional metallic DOS. Assuming spin degeneracy (paramagnetic solutions), this statement can be illustrated by means of the single-particle local spectral function $A_{\mathrm{loc}}(\omega)=\mathrm{Tr} \sum_{\kvec} \boldsymbol{A}(\kvec,\omega)$ with $\boldsymbol{A}(\kvec,\omega)=-\mathrm{Im} \boldsymbol{G}(\kvec,\omega)/\pi$ being the spectral function matrix in the spin/orbital space obtained from the Green's function matrix
\begin{equation}
\boldsymbol{\hat{G}}(\kvec,\omega)=((\omega+\mu -\Sigma(\omega))\mathbf{1}- \boldsymbol{\hat{H}}_0^{(4\times4)}(\kvec) )^{-1}.
\label{eq:Green}
\end{equation}

\begin{figure}[h]
\includegraphics[width=0.5\textwidth]{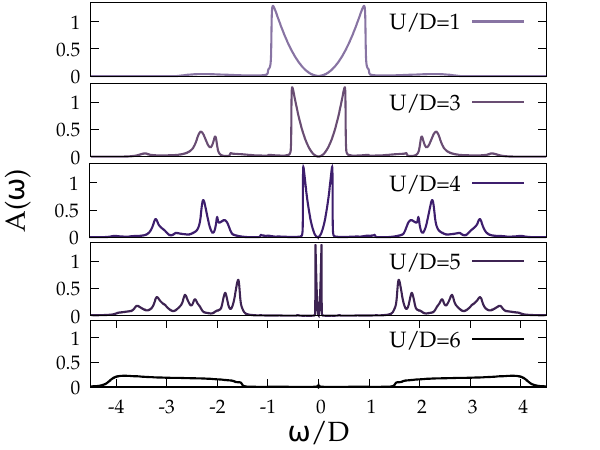}
\centering
\caption{sMIT at half-filling with $\Lambda/D=1$ and $T/D=0.01$ illustrated with the evolution of the local DMFT spectral function. The transition from the semimetallic to the insulating state occurs at a specific critical value $U_{c_2}$ that in this specific case is $U_{c_2}/D=5.6$.
}
\label{fig:MottWeylLambda}
\end{figure}

The evolution of the
semimetallic solution across the sMIT revealed by the IPT solver of DMFT is fairly similar to that of the DMFT solution of the Hubbard model on the cubic or Bethe lattice: the low-frequency central feature in the spectral function, made of a renormalized parabola (henceforth called U-shape),  
loses weight in favour of the upper and lower Hubbard bands at $\omega=\pm\frac{U}{2}$. Such a corresponding U-shape, describing a correlated nodal
semimetal (NSM), finally disappears above $U_{c_2}/D$ (see Fig.~\ref{fig:MottWeylLambda}). 
The width of the parabolic U-shape gets smaller while its curvature increases as the renormalized Fermi velocity decreases. The system remains semimetallic all the way to the sMIT, at which the curvature of the parabolic U-shape diverges and the renormalized Fermi velocity becomes zero. 
At larger interactions, only the Hubbard bands survive. The overall picture is remarkably similar to that found in solving a tight-binding version of Weyl-Hubbard Hamiltonian with exact diagonalization \citep{Topological_Mott_tr_DMFT_WSM_Hubbard}

\begin{figure}[h]
  \begin{tikzpicture}
    \node[inner sep=0pt] (image1) at (0,0) {\includegraphics[width=0.5\textwidth]{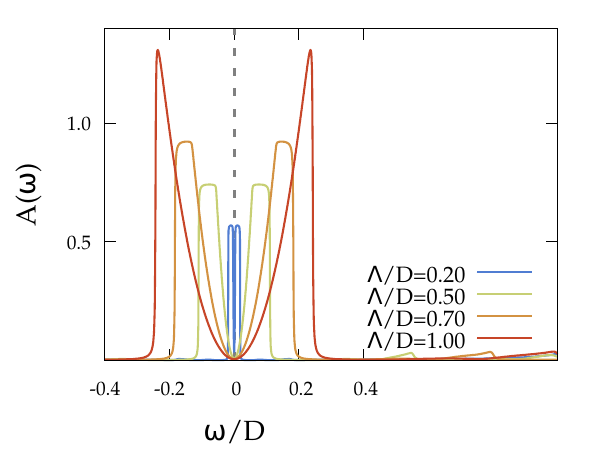}};
    \node[inner sep=0pt] (image2) at (2.1,1.25) {\includegraphics[scale=0.35]{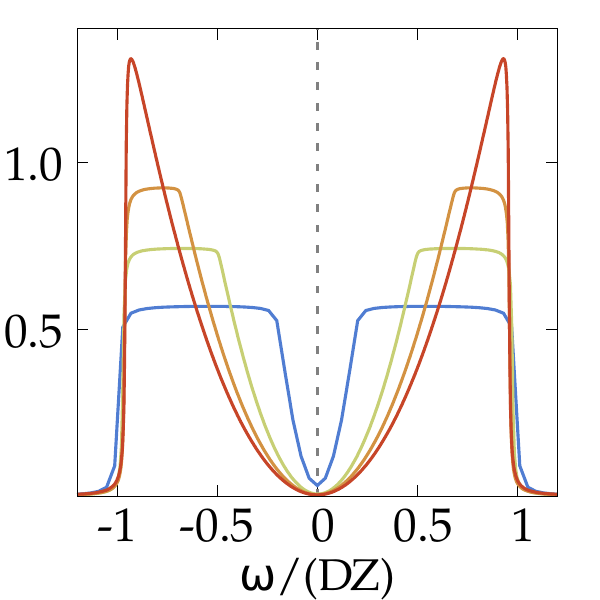}};
  \end{tikzpicture}
\centering
\caption{The low-frequency spectral function at half-filling for $U/D=4.2$ and $T/D=0.01$ in the strongly correlated semimetallic phase for different values of $\Lambda/D$. In the inset, the same quantity with the frequency axis scaled with quasiparticle $Z$. The Fermi level is indicated by the dashed line.
}
\label{fig:AvsLambda}
\end{figure}

Specifically, in Fig. \ref{fig:AvsLambda} is shown the shape of the central coherent part of the DOS at a fixed temperature ($T/D=0.01$) in the strongly correlated NSM phase near the sMIT $U/D=4.2$ for different value of the cutoff $\Lambda/D$.  The overall weight of the central U-shape here shows a strong dependence on $\Lambda/D$ while approaching the sMIT by decreasing $\Lambda/D$ itself (see below). 

In the NSM case, there is no quasi-particle peak at the nodal point therefore the factor $Z$, defined as
\begin{equation}
Z=\left(1-\frac{\partial Re\left[\Sigma(\omega)\right]}{\partial \omega}\right)^{-1}_{\omega=0},
\label{eq:Z}
\end{equation}
should be associated with the renormalization of the Fermi velocity. Indeed, the free particle DOS $A_0(\omega)$ scales quadratically in energy for a linear band structure ($A_0(\omega)\propto \omega^2/v^3_F$) and therefore the self-energy effects included in the renormalization factor $Z$ leads to  $A(\omega)\propto  (\omega/Z)^2/ v^3_F$, which indeed amounts to a rescaling of the Fermi velocity $v_F\rightarrow Zv_F$ as well as to a reduction of the {\it total} spectral weight associated with the central U-shape as is shown in the inset of to Fig. \ref{fig:AvsLambda} where $A(\omega)$ is plotted against $\omega/Z$. 
\begin{figure}[h!]
\begin{tikzpicture}
    \node[inner sep=0pt] (image1) at (0,0) {\includegraphics[width=0.45\textwidth]{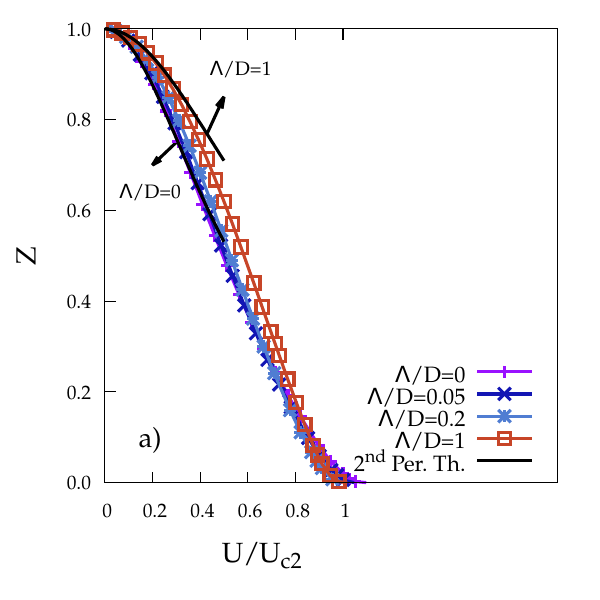}};
    \node[inner sep=0pt] (image2) at (1.8,1.8) {\includegraphics[scale=0.35]{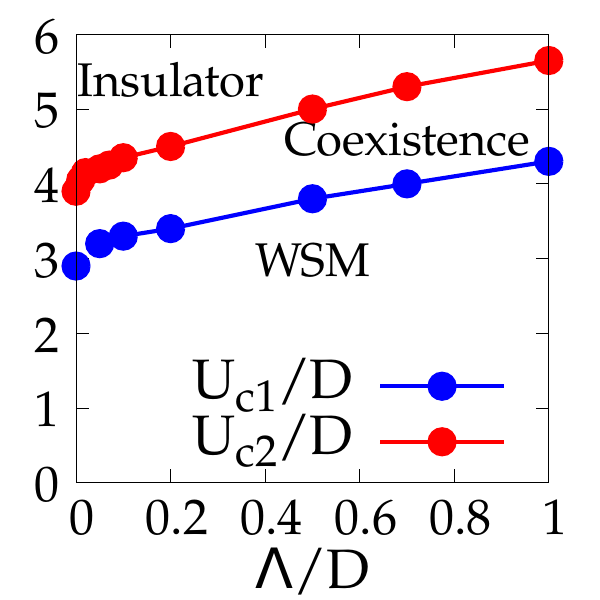}};
\end{tikzpicture} \\
\includegraphics[width=0.45\textwidth]{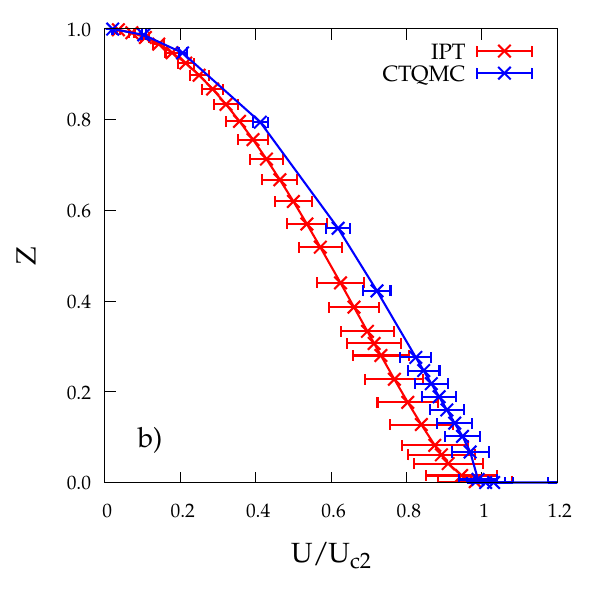}
\centering
\caption{a) The quasiparticle weight Z at half-filling for $T/D=0.01$ is calculated at several values of $\Lambda/D$ as a function of $U/U_{c2}$. The same figure also shows the value obtained for a flat DOS ($\Lambda/D=0$). The black lines represent the second-order diagram calculations both for the Dirac model with $\Lambda/D=1$ \citep{wagner} and the flat one with $\Lambda/D=0$ (Appendix \ref{app:secondordercalc}). In the inset the values $U_{c1}/D$ and $U_{c2}/D$ as a function of $\Lambda/D$ are shown for $T/D=0.01$.
b) Comparison between the renormalization factor $Z$ calculated with IPT and QMC. Horizontal error bars are estimated considering the uncertainty in calculating $U_{c2}/D$, and they are shown in order to specify that the two methods give different values beyond errors but the resulting behaviour is qualitatively similar.
}
\label{fig:ZAllLambda}
\end{figure}
The strong dependence of the factor $Z$ on $\Lambda$ observed in Fig. \ref{fig:AvsLambda} is attributed to the proximity of the sMIT. Interestingly, when plotted against the distance from the sMIT, the factor $Z$ shows little qualitative dependence on $\Lambda$ (see Fig. \ref{fig:ZAllLambda}). Thus, the observed $\Lambda$ dependence of $Z$ primarily arises from the variation of the critical $U$ with $\Lambda$, as illustrated in the inset of Fig. \ref{fig:ZAllLambda} a).

As in the case of the Hubbard model with metallic DOS, the sMIT shows a finite temperature region of coexistence of insulator and NSM solutions between $U_{c_1}$ and $U_{c_2}$, both depending on $\Lambda$. The value of $U_{c1}$ is obtained by starting from the insulating solution and decreasing $U$ until at $U=U_{c1}$ the semimetallic U-shape suddenly appears. 
As for the standard MIT, the IPT solver gives larger critical values $U_{c_1}$ and $U_{c_2}$ when compared with ED \citep{Topological_Mott_tr_DMFT_WSM_Hubbard}. 
In Fig. \ref{fig:ZAllLambda} b) a comparison of IPT and a CTQMC solver \citep{wallerberger_w2dynamics_2019} is shown in the case $\Lambda/D=1.0$ for $T/D=0.01$ as a function of $U/U_{c2}$. The overall behaviour is similar however for this temperature CTQMC predicts $U_{c2}/D=4.85$ while IPT gives $U_{c2}/D=5.6$.

\begin{figure}[h!]
\includegraphics[width=0.45\textwidth]{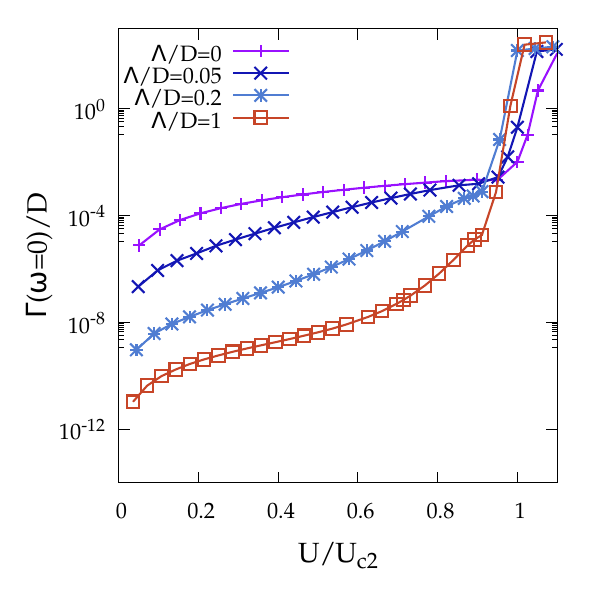}
\centering
\caption{The scattering rate $\Gamma/D$ at half-filling for $T/D=0.01$ calculated at several values of $\Lambda/D$ as a function of $U/U_{c2}$.}
\label{fig:GammaAllLambda}
\end{figure}

Differently from the $Z$ factor, the scattering rate $\Gamma=-\mathrm{Im} \Sigma(\omega=0)$ depends explicitly on the cutoff $\Lambda$ and not only on the distance from the sMIT as is shown in Fig. \ref{fig:GammaAllLambda}. In particular, $\Gamma$ increases by orders of magnitude as far as $\Lambda$ decreases in the NSM phase. We shall discuss the implications of transport of this behaviour in the next section.

\begin{figure}[h]
\begin{tikzpicture}
    \node[inner sep=0pt] (image1) at (0,0) {\includegraphics[width=0.5\textwidth]{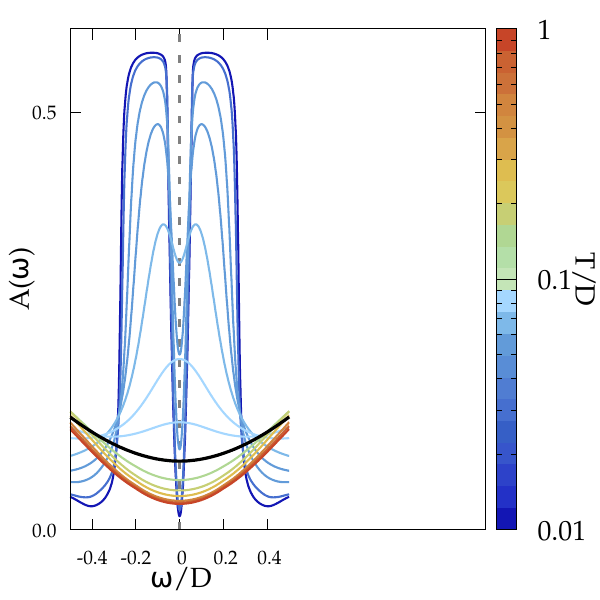}};
    \node[inner sep=0pt] (image2) at (1.1,2.3) {\includegraphics[scale=0.37]{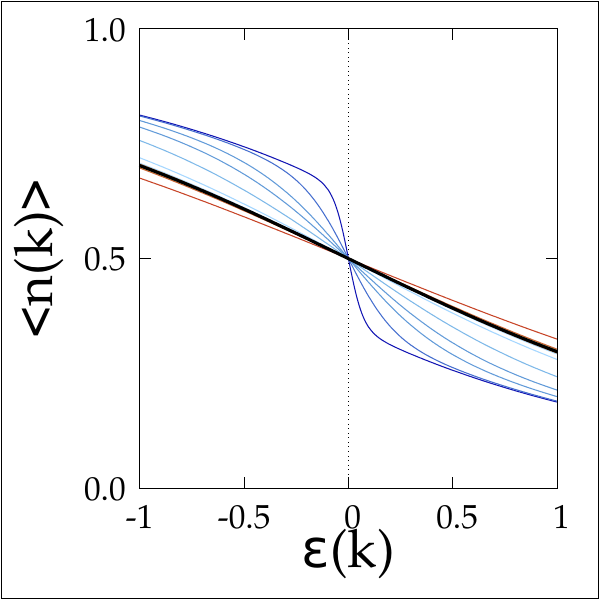}};
\end{tikzpicture}
\centering
\caption{ 
The central U-shape in the spectral function at half-filling for $U/D=3.0$ and $\Lambda/D=0.2$ as a function of temperatures. In the inset, the momentum distribution as a function of $\epsilon(\kvec)$. The black bold line is the $T^*/D=0.1$. The vertical dashed line indicates the position of the Fermi level.
}
\label{fig:AvsTemp}
\end{figure}

In Fig. \ref{fig:AvsTemp} the evolution of the central U-shape is shown for increasing temperature at the given value of $U/D=3$ in the strongly correlated NSM phase for $\Lambda/D=0.2$. This low-energy spectral structure disappears quickly when increasing the temperature, going from the coherent semimetallic state to the incoherent bad metal. This is also reflected in the momentum distribution ($\langle n(\kvec) \rangle=\int d\omega \mathrm{Tr} \boldsymbol{A}(\kvec,\omega)$) (inset of Fig. \ref{fig:AvsTemp}), where the sharp decrease around $\epsilon(k)=0$ signalling the semimetallic Fermi point around $k=0$ progressively disappears as the temperature reaches the scale $(T/D)^*\sim 0.1$ where the semimetallic peak disappears (black curves in Fig. \ref{fig:AvsTemp}). 
In the DMFT approach to the MIT in the single band Hubbard model the temperature above which  the spectral function becomes pseudogapped has been interpreted as the Kondo scale of the associated Anderson impurity model (AIM) \citep{georges_dynamical_1996,chalupa_localmoments_PRL2021,AndreyPhysRevB.105.L081111}.

The fact that a Kondo screening is observed in the NSM strongly correlated phase despite the vanishing of the DOS at the Fermi point can indeed be rationalised by looking at the peculiar nature of the associated AIM.
As in the standard single site single-orbital DMFT, in our case, the local Green function (see Appendix \ref{app:DMFT}) can be expressed as $\hat{G}=(\omega+\mu -\Delta-\Sigma)^{-1}\mathbf{1}$ in terms of the hybridisation function $\Delta$ of an AIM. If one assumes as customary for non-pseudo-gapped system that $-Im\Delta(\omega) \propto -Im G(\omega)$ for $\omega\simeq 0$ then one should not expect Kondo screening to occur because the lack of states around the nodal point \citep{VojtaKondoPseudogapPRB2004}. However, the semimetallic solution shows instead a {\it diverging} $\Delta$ around the zero energy since to get a vanishing DOS i.e. $-Im G(\omega=0)=0$ one needs a diverging $Im \Delta$ assuming that the self-energy is still irrelevant in the semimetallic phase around the nodal point. This large hybridisation guarantees the presence of Kondo screening even when the DOS vanishes.

\section{Transport}
\label{sec:transport}

d.c. and a.c. conductivity are obtained using the Kubo formula within the standard DMFT approach which neglects vertex corrections (see Appendices \ref{app:Kubo} and \ref{app:optcond}). Here we exploit the Nernst-Einstein relation
\begin{equation}
    \sigma=e^2 \chi \mathcal{D}
\label{eq:Nernst}
\end{equation}
where $\chi$ is the charge compressibility
\begin{equation}
\chi=\frac{\partial \langle n \rangle}{\partial \mu}
\label{eq:compressibility}
\end{equation}
and $\mathcal{D}$ is the charge diffusivity. In a previous work  \citep{wagner} some of us derived the asymptotic behaviour for the conductivity at the nodal point as well as for finite chemical potential in the semimetallic regime establishing an approximate Drude-Boltzmann approximation for linearly dispersing bands
\begin{equation}
    \sigma=\frac{e^2n_{eff}\tau}{m^*}
\label{eq:drude}
\end{equation}
where $n_{eff}/m^*$ is the effective number of carriers (scaled with effective mass) and $\tau$ is a scattering time   defined in terms of quasiparticle scattering rate $\Gamma$.
At {\it weak coupling} one can easily prove that $n_{eff}\propto \chi$ and $\tau \propto \mathcal{D}$ \citep{Ferrero_PhysRevB.107.155140}. 
We here consider the outcomes of the DMFT process away from half-filling by evaluating numerically $\chi$ as the derivative of $\langle n \rangle$ over $\mu$ and computing the diffusivity spanning from weak to strong coupling. In particular, we will identify different crossovers in terms of different temperature exponents $\alpha$ of the quantities appearing in Eq. (\ref{eq:Nernst}) which are related by $\alpha_{\sigma}=\alpha_{\chi}+\alpha_{\mathcal{D}}$.

\subsection{Temperature crossover at nodal point}

A study of the three transport quantities mentioned before ($\sigma$, $\chi$, and $\mathcal{D}$) has been performed as a function of the temperature at half-filling where the Fermi energy is chosen to be exactly at the nodal point. 

It is worth remembering here some results valid in the single band Hubbard model at half-filling for the Bethe model. At weak coupling, in the Fermi Liquid state, the system behaves as a conventional metal therefore, in the Drude approach, the temperature dependence of the conductivity arises entirely from the scattering time. In the Nernst-Einstein approach indeed the compressibility is almost constant w.r.t. both temperature and chemical potential assuming that $T \ll D$ and $|\mu| \ll D$. The temperature dependence of the conductivity is therefore solely due to the temperature dependence of the diffusivity i.e. $\mathcal{D} \propto \tau \propto T^{-2}$ and in terms of temperature exponents $\alpha_{\sigma}=\alpha_{\mathcal{D}}=-2$.

In the opposite limit of incoherent scattering when  $T \gg T^*$, the  linear behaviour for the resistivity is solely due to the temperature dependence of the compressibility, while the diffusion constant is almost constant, i.e. $\alpha_{\sigma}=\alpha_{\chi}=-1$ 
\citep{pakhira_absence_2015}. 

The analysis of Ref. \citep{wagner} for the case of pure quadratic DOS ($\Lambda/D=1.0$) at half-filling can be then reinterpreted in terms of temperature exponents for $\sigma$,$\chi$ and $\mathcal{D}$. The vanishing of the spectral function at the nodal point induces qualitative deviations from the conventional Fermi liquid behaviour in the weak coupling regime.
In this case both $\chi$ and $\mathcal{D}$ contribute to the temperature exponent of conductivity, namely $\alpha_{\chi}=2$ while $\alpha_{\mathcal{D}}=-8$ that corresponds to  $\alpha_{\sigma}=-6$ \citep{wagner}. From the point of view of the Nernst relation, the large diffusivity of the states near the nodal point overcompensates the vanishing charge susceptibility i.e. the vanishing number of available carriers in Eq. (\ref{eq:drude}) leading to a large temperature exponent for resistivity. Moving away from half-filling ($|\mu|<T$), the low-temperature behaviour becomes dominated by the presence of a finite spectral function at the chemical potential and Fermi-liquid exponents are expected $\alpha_{\sigma}=\alpha_{\mathcal{D}}=-2$.

\begin{figure}[h]
\begin{minipage}[t]{0.49\textwidth}
\includegraphics[width=.33\textwidth]{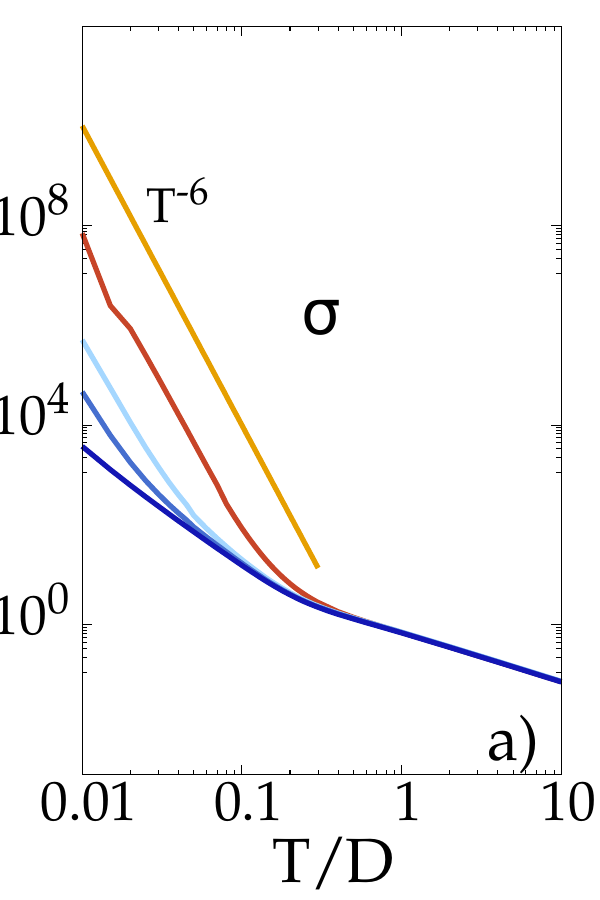}\hspace{-0.1cm}
\includegraphics[width=.33\textwidth]{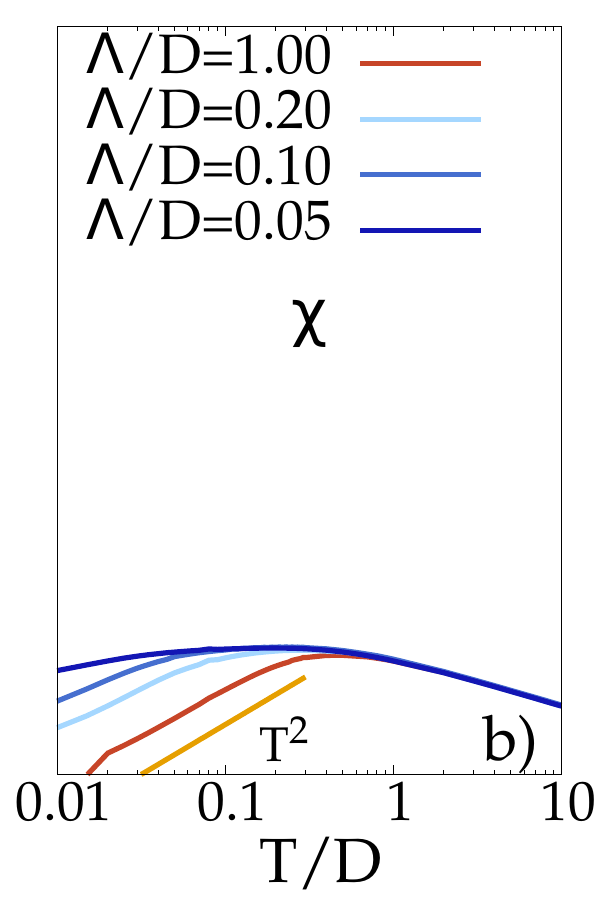}\hspace{-0.1cm}
\includegraphics[width=.33\textwidth]{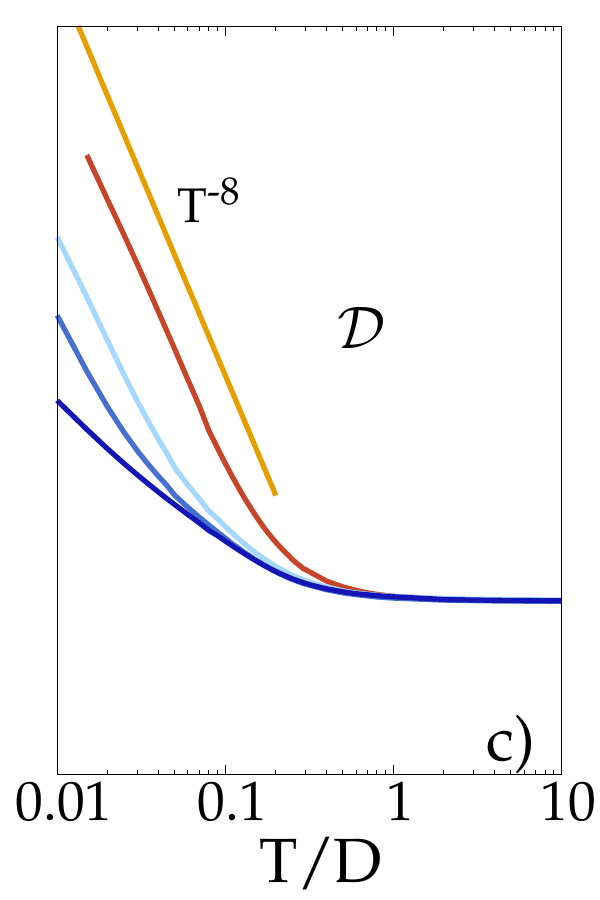}
\hfill
\includegraphics[width=.33\textwidth]{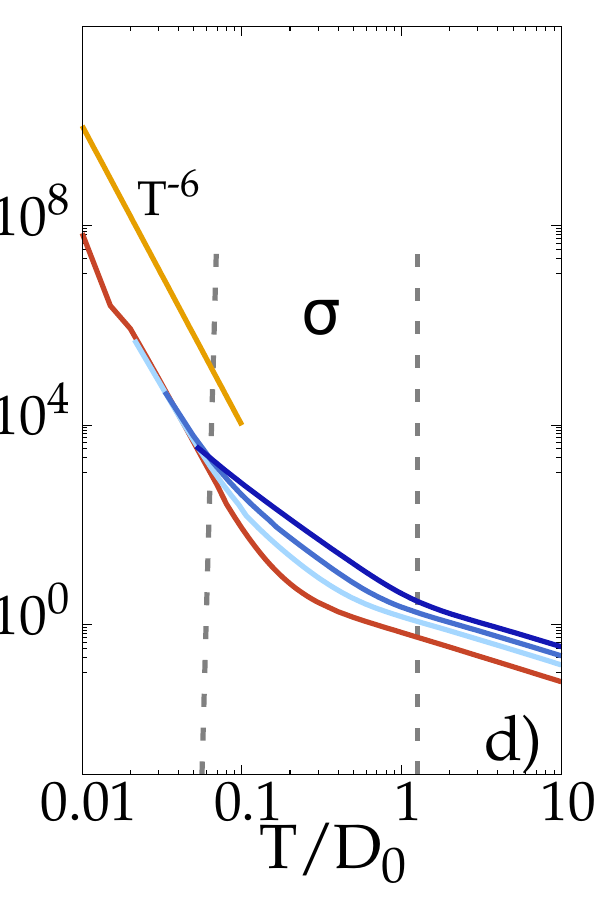}\hspace{-0.1cm}
\includegraphics[width=.33\textwidth]{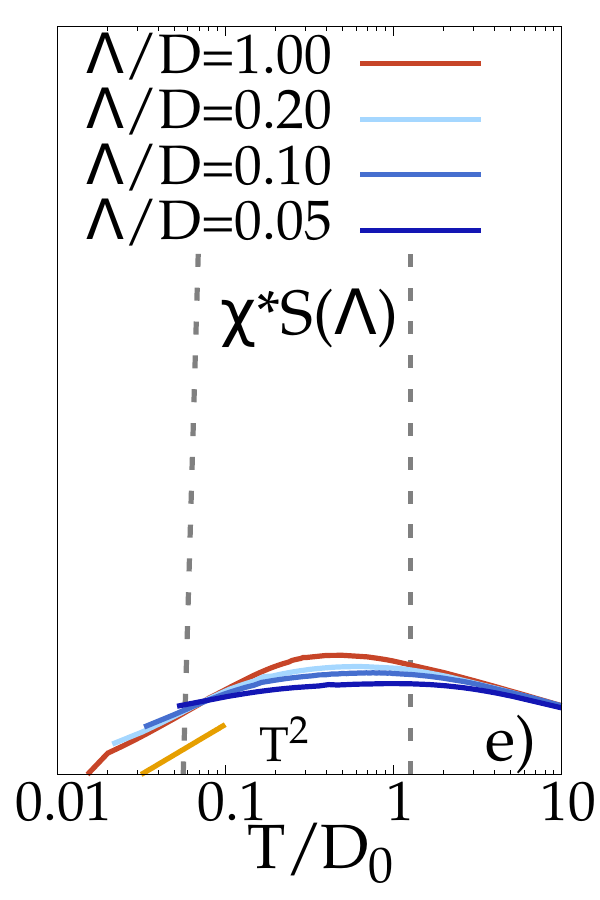}\hspace{-0.1cm}
\includegraphics[width=.33\textwidth]{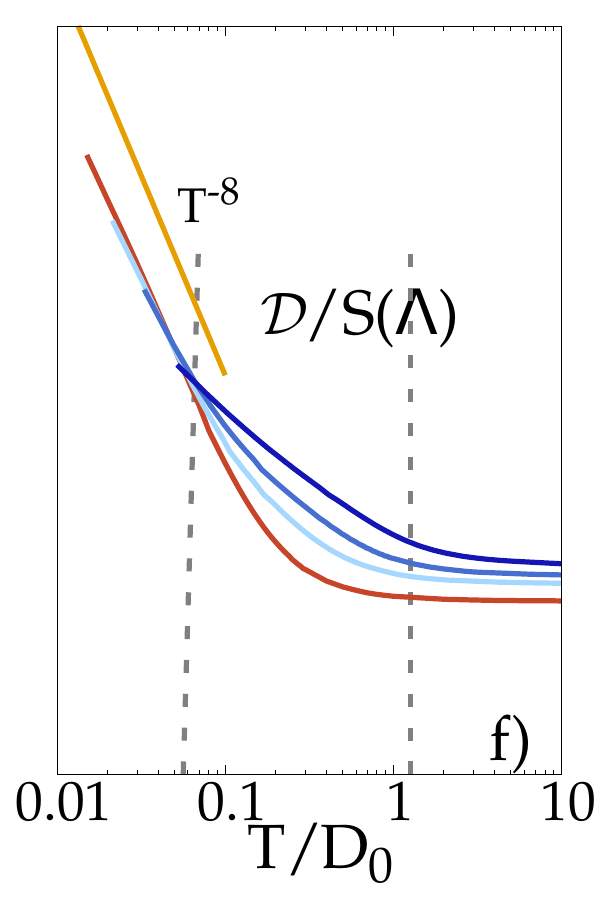}
\centering
\caption{ Conductivity (a), charge compressibility (b) and diffusivity (c) at $\mu=0$ as a function of the temperature at weak coupling ($U/D=0.5$). The same quantities (respectively d), e) and f)) at $\mu=0$ as a function of the scaled temperature at weak coupling ($U/D=0.5$). The dashed grey lines are useful to separate the crossover central region from the NSM one at lower temperatures and the bad metal one at higher temperatures.}
\label{fig:Rho_vs_T}
\end{minipage}
\end{figure}

These exponents are further modified by the presence of cut-off energy $\Lambda$ leading to temperature and density crossover.
Evidently, in Fig. \ref{fig:Rho_vs_T} we show the conductivity (a), charge compressibility (b) and diffusivity (c) at the nodal point as a function of the temperature at weak coupling ($U/D=0.5$). 
The low-energy behaviour can be studied by introducing a scaling method. This scaling method is designed to bring different models with different values of the cutoff $\Lambda/D$ to the same Fermi velocity, thereby achieving the same curvature of the density of states around zero energy, as described in Appendix \ref{app:scaling}. In Fig. \ref{fig:Rho_vs_T} (d), e) and f)) we present the same coefficients as shown in Fig. \ref{fig:Rho_vs_T} (a), b) and c)) but plotted as a function of scaled temperatures at weak coupling ($U/D=0.5$). The scaling approach enables a collapse of the curves at low temperatures for various values of the cutoff  $\Lambda/D$.

As explicitly shown in Appendix \ref{app:scaling}, this scaling introduces a factor $S(\lambda)={\left(\lambda^3+3\lambda^2(1-\lambda)\right)^{\frac{1}{3}}}$ where $\lambda=\Lambda/D$, and the temperature is calculated in units of a fixed half-bandwidth $D_0=S(\lambda)D$.
The scaled picture highlights that the limiting temperature behaviour found for $T/D_0\ll1$ shows the characteristic low-temperature exponents $\alpha_{\sigma}=-6$, $\alpha_{\chi}=2$, $\alpha_{\mathcal{D}}=-8$. However, by increasing the temperature we enter the intermediate temperature {\it metallic} phase characterized by the same exponents found for a Fermi liquid i.e. $\alpha_{\sigma}-2$, $\alpha_{\chi}=0$ and $\alpha_{\mathcal{D}}=-2$.
By further increasing the temperature, when $T/D>1$ the incoherent $\Lambda$-independent regime is reached. This regime is clearly seen in Fig. \ref{fig:Rho_vs_T} where $\alpha_{\sigma}=-1$, $\alpha_{\chi}=-1$, $\alpha_{\mathcal{D}}=0$, i.e. the same as for the incoherent bad metallic regime \citep{pakhira_absence_2015}.

\begin{figure}[h]
\begin{minipage}[t]{0.49\textwidth}
\includegraphics[width=.33\textwidth]{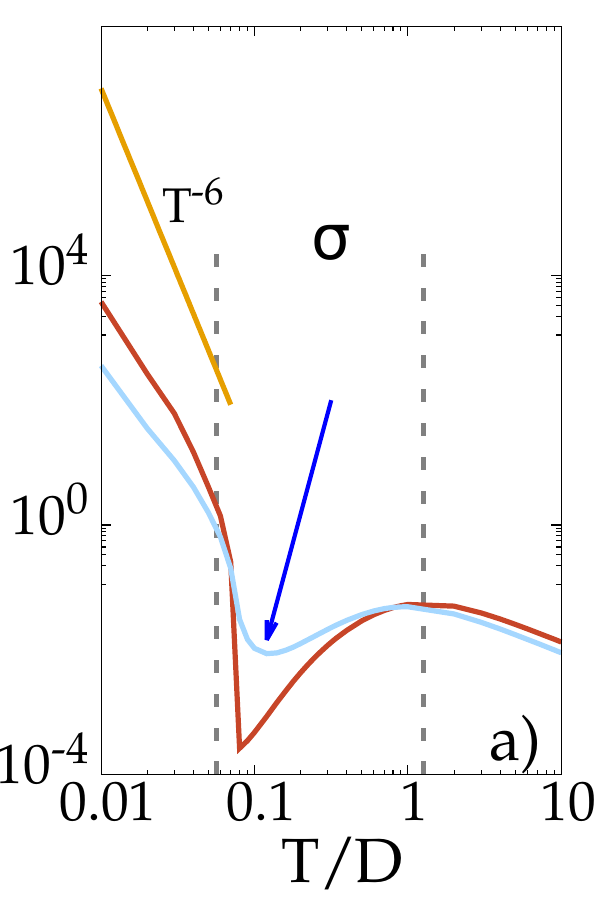}\hspace{-0.1cm}
\includegraphics[width=.33\textwidth]{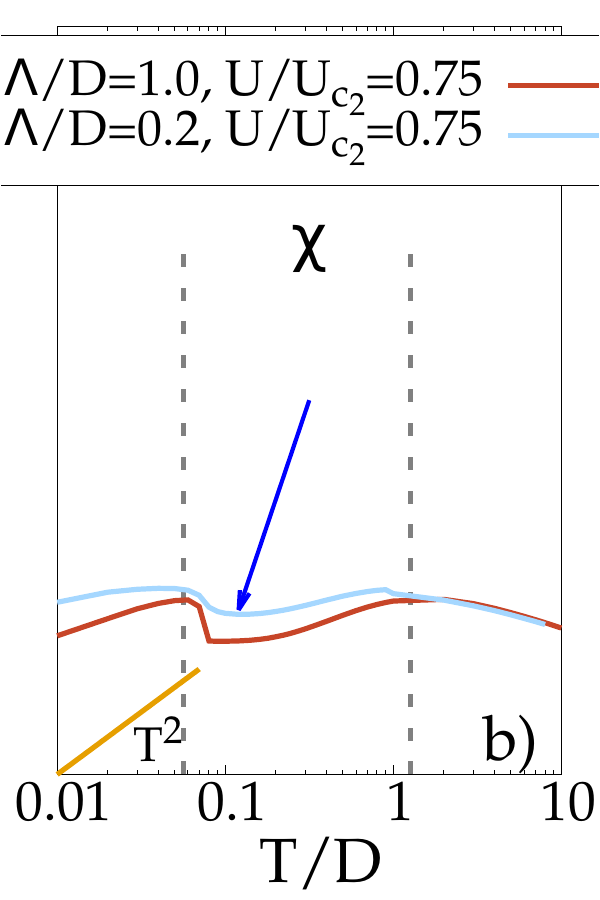}\hspace{-0.1cm}
\includegraphics[width=.33\textwidth]{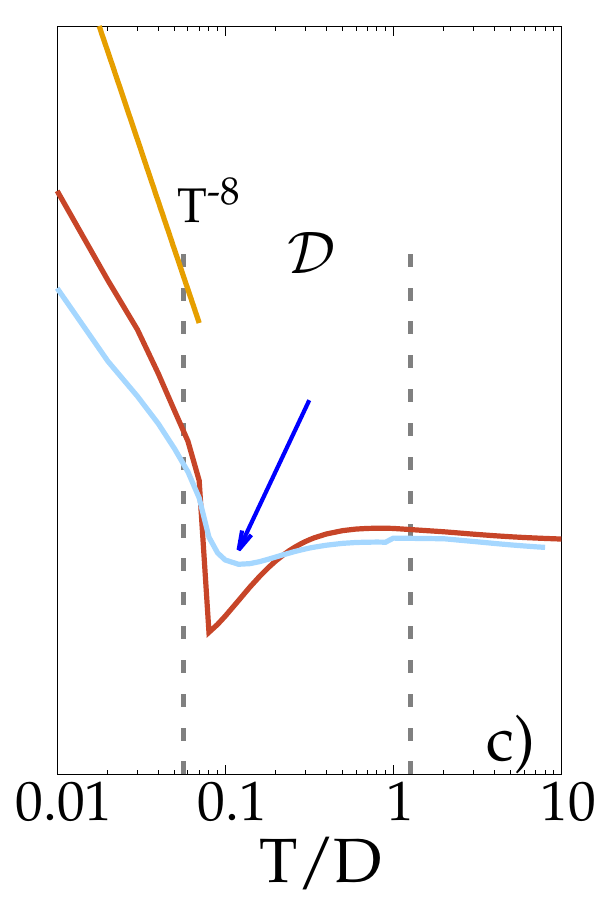}
\centering
\caption{Conductivity (a), charge compressibility (b) and diffusivity (c) at $\mu=0$ as a function of the temperature at strong coupling near the sMIT.
Arrows indicate the temperature at which the quasiparticle U-shape disappears ($T^*/D=0.1$). The dashed grey lines are useful to separate the transition central region from the NSM one at lower temperatures and the bad metal one at higher temperatures.
}
\label{fig:NernstStrongCoupling}
\end{minipage}
\end{figure}

In Fig. \ref{fig:NernstStrongCoupling} the quantities $\sigma$, $\chi$ and $D$ are shown in the strongly correlated NSM phase ($U/U_{c_2}=0.75$) at half-filling. At low temperatures  ($T<T^*$)  a semimetallic behaviour is observed but with apparent exponents which slightly deviate from that of the weak coupling at $\Lambda/D=1$ as shown in the figure. Deviation from the exponents observed in the weak-coupling regime gets more marked for $\Lambda/D<1$. 
By increasing the temperature we eventually meet the first-order finite temperature transition for $\Lambda/D=1$ which appears to be smoothed out for smaller $\Lambda$. 
By further increasing the temperature a non-monotonic behaviour of $\sigma$ and $\chi$ is observed after the disappearance of the coherent central U-shape at $T^*/D$ (arrows in Fig. \ref{fig:NernstStrongCoupling}). 
A final bad metallic regime with the conductivity and compressibility decreasing linearly with $T$ is reached on a temperature scale larger than the bare half-bandwidth. 
Interestingly the sharp variation of charge compressibility observed around $T=T^*$ in Fig. \ref{fig:NernstStrongCoupling} b) is similar to that found in the strongly correlated metallic phase of the single-band Hubbard model \citep{pakhira_absence_2015}. 
There, the charge susceptibility increases upon decreasing the temperature and reaches a constant value at $T/D=0$ due to the onset of Kondo screening of the local moments.  In contrast, here, $\chi$ after a first increase finally decreases when lowering the temperature as expected from the vanishing of the spectral function. However, its initial growth which occurs as far as the central U-shape is restored for $T<T^*$ indirectly suggests that an interpretation of the results in terms of the screening of local moments is possible in the strongly correlated NSM at least on an intermediate temperature scale due to the finite, albeit small, spectral weight at the nodal point \citep{georges_dynamical_1996,chalupa_localmoments_PRL2021}.

\subsection{Density crossover away from nodal point}

\begin{figure}[h]
\begin{minipage}[t]{0.49\textwidth}
\includegraphics[width=.3\textwidth]{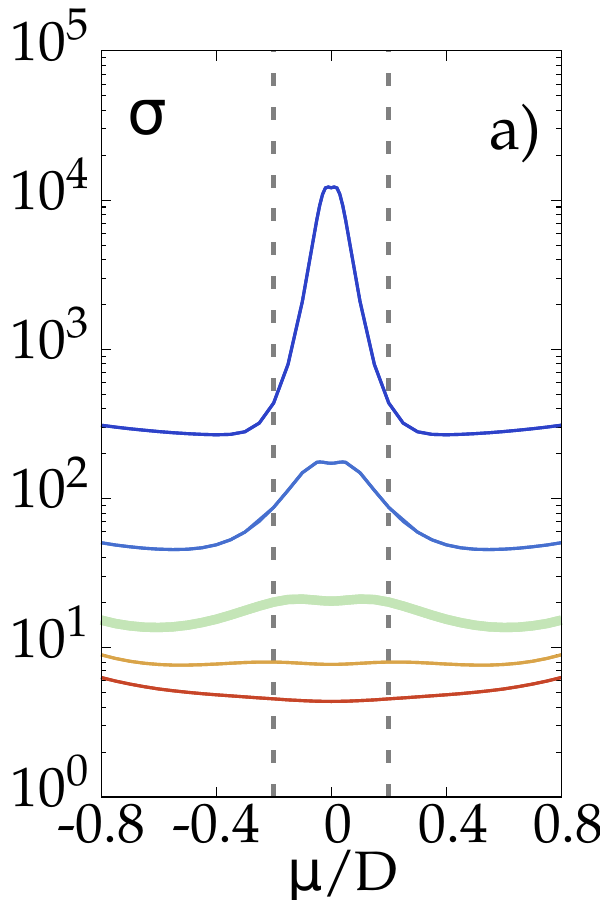}\hspace{-0.1cm}
\includegraphics[width=.3\textwidth]{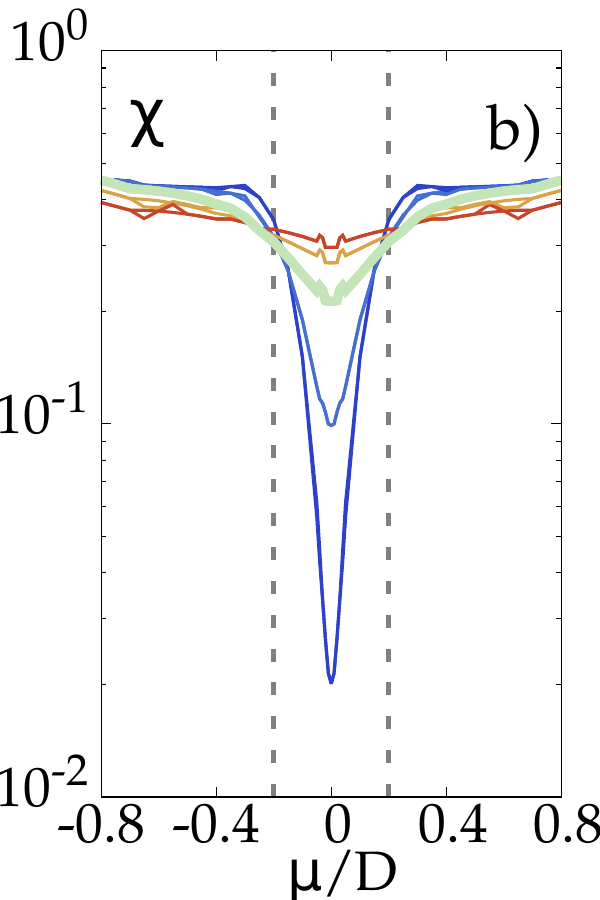}\hspace{-0.1cm}
\includegraphics[width=.3\textwidth]{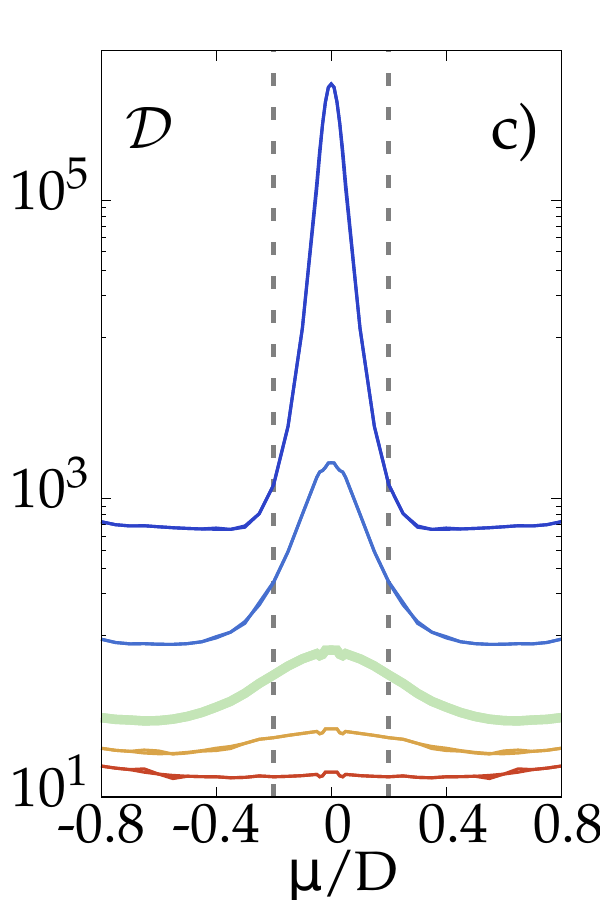}\hfill
\includegraphics[scale=0.13]{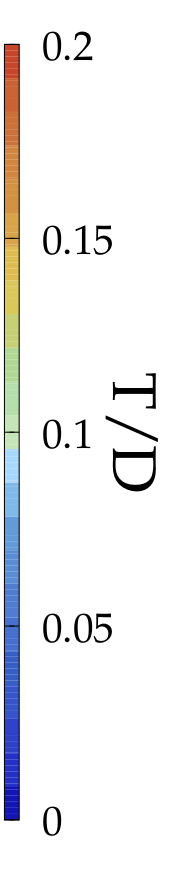}
\centering
\caption{Conductivity (a), charge compressibility (b) and diffusivity (c) for $U/D=0.5$ and $\Lambda/D=0.2$ as a function of the chemical potential. Different curves refer to different temperatures. Dashed grey lines indicate the width of the U-shaped structure in the spectral function.
}
\label{fig:Nernstmu_WeakCoupling}
\end{minipage}
\hfill
\begin{minipage}[t]{0.49\textwidth}
\includegraphics[width=.3\textwidth]{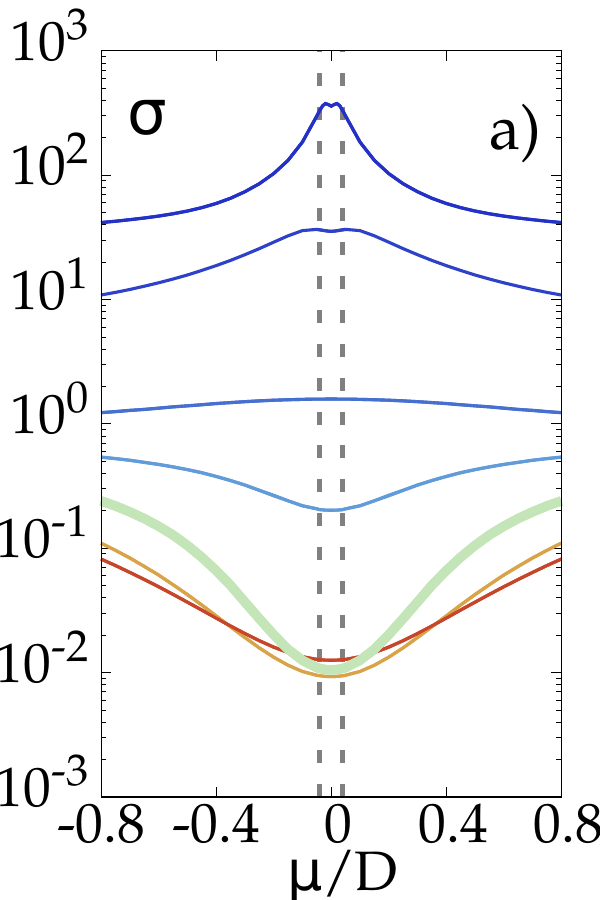}\hspace{-0.1cm}
\includegraphics[width=.3\textwidth]{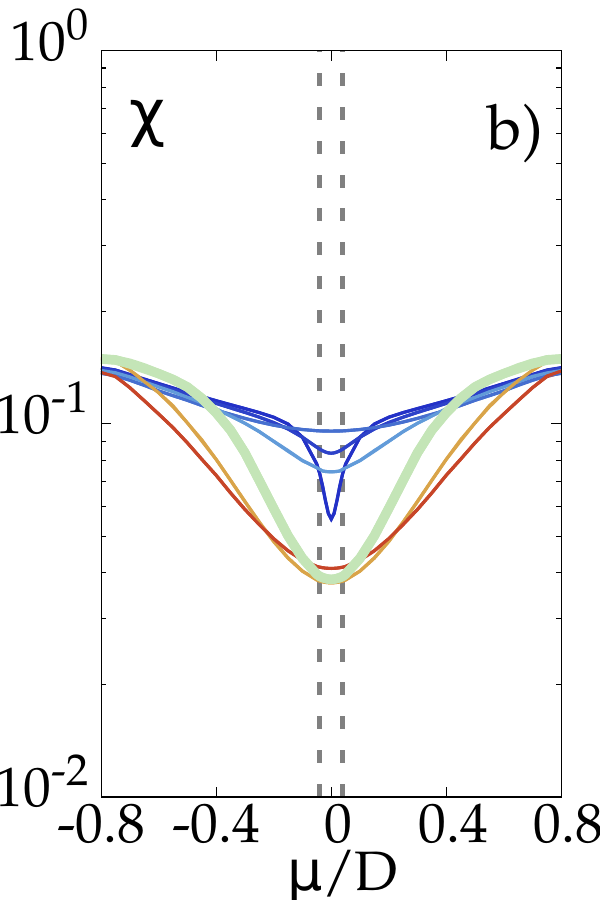}\hspace{-0.1cm}
\includegraphics[width=.3\textwidth]{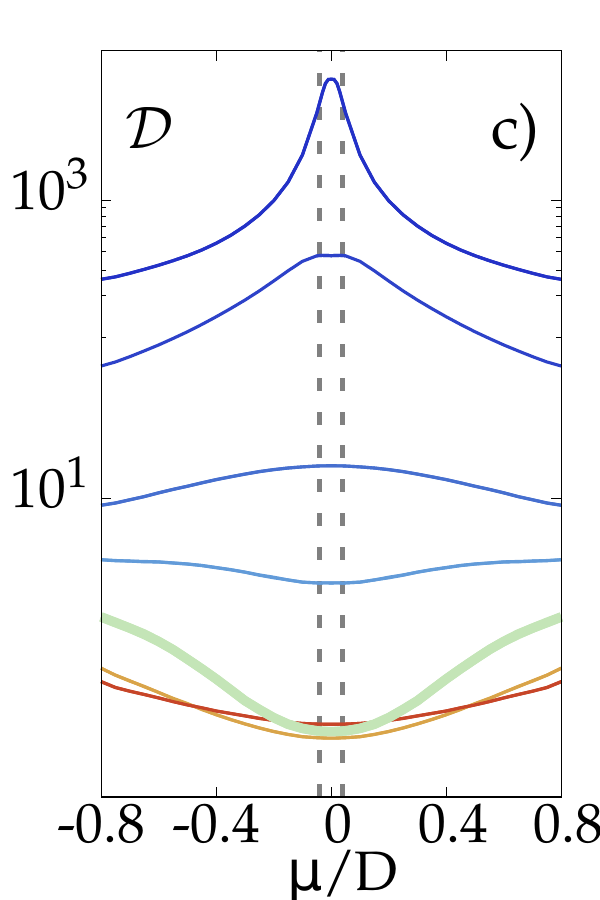}\hfill
\includegraphics[scale=0.13]{Tscale.png}
\centering
\caption{Conductivity (a), charge compressibility (b) and diffusivity (c) for $U/D=3.0$ and $\Lambda/D=0.2$ as a function of the chemical potential. Different curves refer to different temperatures. The green bold line ($T^*/D=0.1$) marks the temperature at which, at half-filling, the quasiparticle U-shape vanishes. Dashed grey lines indicate the width of the U-shaped structure in the spectral function.
}
\label{fig:Nernstmu_StrongCoupling}
\end{minipage}
\end{figure}

Figures \ref{fig:Nernstmu_WeakCoupling} and \ref{fig:Nernstmu_StrongCoupling} show the quantities involved in the Nernst relation as a function of the chemical potential for different temperatures. While conductivity is obtained by the Kubo formula Eq. (\ref{eq:Kubo}), the compressibility has been obtained by probing the electron density as a function of chemical potential for constant temperatures as shown in Appendix \ref{app:nvsmu}.

At weak coupling and low temperatures ($T<\Lambda$)  (Fig. \ref{fig:Nernstmu_WeakCoupling}) we can identify tree regions: 

i) When $|\mu|\ll T$,  the system is really close to the nodal point and the temperature exponents are those found at the nodal point \citep{wagner}. In this region, thermal fluctuations probe energy ranges in which the spectral density is very small and the behaviour of all quantities tends to be weakly dependent on $\mu$.

ii) When  $T<\mu<\Lambda$ the behaviour in $\mu$ is that of a NSM with linear bands: the diffusivity decreases and the compressibility increases with $\mu$ as expected from the increasing of the spectral density. The smaller the temperature the sharper the conductivity drop. 

iii) When $\mu>\Lambda$ the transport properties have a Fermi liquid behaviour dominated by the non-linear part of the band structure. 

A clear distinction between these zones disappears gradually as $T$ approaches $\Lambda$ when the semimetallic regime is gradually washed out.

In the strongly correlated NSM phase, (Fig. \ref{fig:Nernstmu_StrongCoupling}) the very low-temperature $\mu$-dependence of all quantities resembles qualitatively that of the weakly interacting case only at very low temperatures. 
One may note that with increasing $U$, the energy range where the quadratic pseudogap is present is narrower (Fig. \ref{fig:AvsTemp}) and therefore the crossover region ii) seen at weak coupling is here restricted to very low temperature (here $T/D<0.05$).
The strong reduction of the spectral weight at the minimum of the U-shape quasiparticle feature  (see Fig. \ref{fig:AvsTemp})
occurring around $T\simeq T^*$ at $\mu=0$ marks here a qualitative change of the behaviour in $\mu$ of all quantities: for $T>T^*$ $\sigma$ increases with increasing chemical potential as a result of the concomitant enhancement of both $\chi$ and $\mathcal{D}$ which can here be ascribed to the partial restoring of the coherent U-shape with increasing $\mu$ (see Appendix \ref{app:Filling}).

\subsection{Energy crossover at nodal point and optical damping}

\begin{figure}[h]
\includegraphics[width=0.35\textwidth]{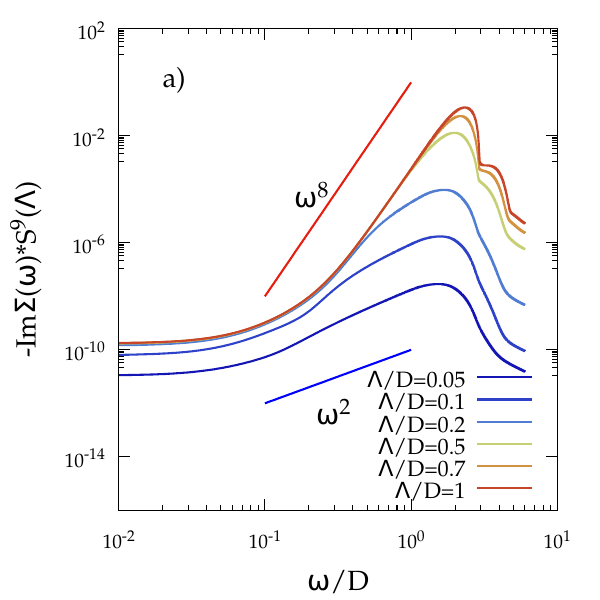}
\includegraphics[width=0.35\textwidth]{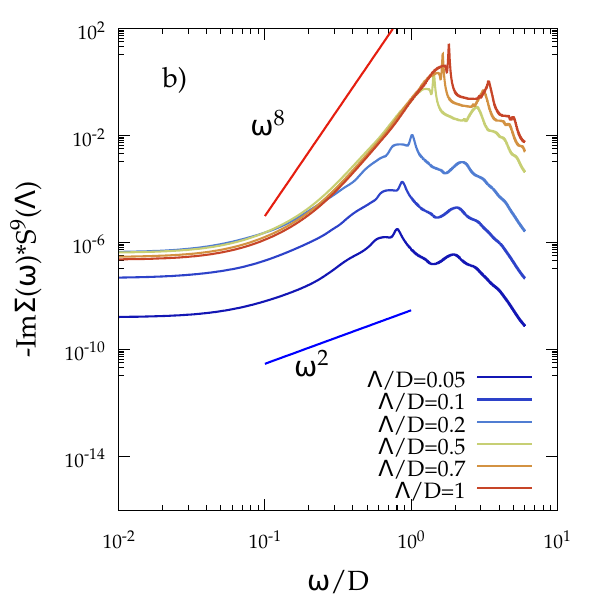}
\centering
\caption{The imaginary part of the self-energy at half-filling for $U/D=0.5$ and $T=0.02$ scaled with  $S^9(\Lambda)$ in order to obtain the same Fermi velocity of any $\Lambda/D$ (a). The same quantity for $U/D=3.0$  (b).
}
\label{fig:Im_Sigma}
\end{figure}

The $\Lambda$-dependent crossover seen at the nodal point in the weak coupling regime (Fig. \ref{fig:Rho_vs_T}) has a counterpart in an energy crossover for the frequency-dependent scattering rate shown in Fig. \ref{fig:Im_Sigma} at constant temperature $T/D=0.02$ and scaled with the scaling factor $S^9(\Lambda)$ to compare systems with the same Fermi velocity $v_F$. The scaling dimensions $S^9(\Lambda)$ can be recovered from the dependence on $v_F$ of $\Gamma$, see Eq. (\ref{eq:NSMImaginarySelf}).
The predicted polynomial behaviour  of the scattering rate at nodal point \citep{wagner} spanning from a constant at low energy to $\omega^8$ at higher energies is shown here for $\Lambda/D=1$ (see also Appendix \ref{app:damping}). For $\Lambda/D<1$ a crossover can be seen at $\omega\simeq \Lambda$, while above this energy scale, a Fermi liquid $\omega^2$-behaviour is recovered. 

In panel b) the scattering rate is shown in the strong coupling NSM phase. For $\Lambda/D=1$ the energy exponent again spans from 0 to 8.

For $\Lambda/D<1$ a crossover to smaller energy exponents is found at $\omega\simeq \Lambda$ though less evident than in the weak coupling case. When the energy scale $\Lambda$ approaches the order of the temperature, the low-temperature scaling fails (see Appendix \ref{app:scaling}).

\begin{figure}[h]
\includegraphics[width=0.39\textwidth]{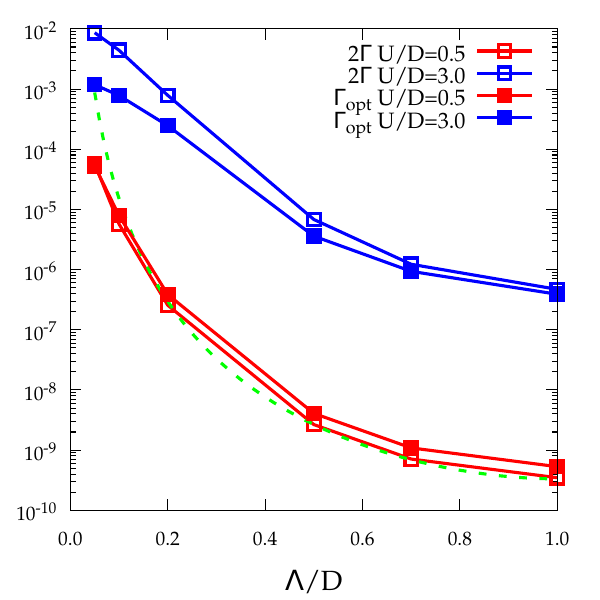}
\centering
\caption{$\Gamma_{opt}$ and $\Gamma$ at half-filling for $T/D=0.02$ and $U/D=0.5$ and $U/D=3.0$. The dashed green line represents the $\Lambda/D$ dependence from the perturbation theory (Eq. \ref{eq:NSMImaginarySelf}). 
}
\label{fig:OptGamma}
\end{figure}

The optical absorption is the result of both intraband and interband contributions \citep{tabert_optical_2016}. Appendix \ref{app:optcond} shows sample data of optical spectra. Here, we focus our attention on the low energy optical spectra which is due to the intraband transition and has a Drude form
\begin{equation}
    \sigma(\Omega)=\frac{\sigma(0) \Gamma^2_{opt}}{\Omega^2+\Gamma^2_{opt}}    
\end{equation}
The low energy peak is very narrow due to the smallness of the scattering rate around the nodal point. As it is shown in Fig. \ref{fig:OptGamma} the optical scattering rate at weak coupling does depend strongly on the amplitude of the energy scale $\Lambda$ through the same scaling function introduced in Fig. \ref{fig:Rho_vs_T} since $\Gamma_{opt}\simeq 2\Gamma$ (Appendix \ref{app:optcond}). Scaling does not hold in the NSM strongly correlated phase 
where $\Gamma_{opt}$ is not simply proportional to the scattering rate as shown in Fig. \ref{fig:OptGamma}. Although the qualitative dependence on $\Lambda$ is similar in both cases, we notice a significative deviation of the optical scattering rate from its estimate based on the quasiparticle scattering rate when non-linear deviations become important.

\section*{Conclusions}

In this work, we have studied the correlation effects on a Dirac-semimetal which leads to a semimetal-to-insulator transition (sMIT) at large values of U. Compared to the standard description of MIT within DMFT, no quasiparticle peak appears. Instead the semimetallic U-shape
peak is preserved up to the transition to the Mott insulator. Its shape is renormalized and its width decreases with increasing interaction strength while spectral weight is transferred to the Hubbard bands. This coherent low-temperature phase, as in the case of DMFT calculations for the Bethe lattice, disappears upon increasing the temperature.

While the overall features of the sMIT are not much affected by the presence of an energy scale $\Lambda$ which separates the linear from a non-linear momentum dependence in the band structure, the presence $\Lambda$ largely affects transport properties. Several crossovers are found to be crucially dependent on the presence of such a scale with the analysis of the Nernst relation involving the conductivity, the compressibility and the diffusivity with respect to the temperature, the chemical potential and the energy in the weakly correlated phase. 

At the nodal point, a crossover in temperature is found toward a Fermi-liquid-like behaviour as the temperature exceeds a $\Lambda$-dependent scale. In the same way, at a given temperature if the chemical potential exceeds the $\Lambda$ scale we enter into a Fermi-liquid-like regime.

The energy-dependent scattering rate shows a similar crossover where the NSM character dominates at low energy and is progressively replaced by a Fermi-liquid-like behaviour at higher energies. This is reflected in a strong $\Lambda$ dependence on the optical scattering rate.

By increasing electronic correlations a coherent behaviour which shows the previously mentioned crossovers does exist only for very low temperatures. Interestingly, we also found an intermediate temperature window in which the behaviour of the charge compressibility may be indicative of a partial screening of the local moments. Ongoing work specifically dedicated to this aspect will help to clarify the precise role played by Kondo screening processes in NSM \citep{MaxInPreparation}.

Another point which deserves further investigation which is beyond the scope of the present work is the role of the vertex corrections on the transport properties. Vertex corrections may introduce different temperature dependence in transport and quasiparticle scattering as it has been shown for Dirac fermions in two dimensions \citep{lara_bft_2009} or may lead to deviations in low-frequency optical conductivity at low-density \citep{MillisAnqiPRB2022}.

\begin{acknowledgments}
G.S. acknowlegde useful discussion with Björn Trauzettel.
S.C. acknowledges useful discussions with Serge Florens. 
S.C. acknowledges funding from  NextGenerationEU National Innovation Ecosystem grant ECS00000041 - VITALITY - CUP E13C22001060006. A.T. acknowledges financial support from the Austrian Science Fund  (FWF), within the project  I 5868  (Project P01, part of the FOR 5249 [QUAST] of the German Science Foundation, DFG)

\end{acknowledgments}

\bibliography{references.bib}

\appendix
\section{DMFT approach to Nodal semimetal}
\label{app:DMFT}

The purpose of this appendix is to introduce a simplified model for the density of states (DOS) being the relevant quantity which enters in the calculations of the spectral functions in the DMFT approach and in the calculation of the conductivity from the Kubo formula. Our model {\it total} density of states i.e. the DOS of all orbital and spin degree of freedom described in the main text (Introduction) is characterized by an intermediate cutoff in energy ($\Lambda$) separating the low-energy part (linear bands) and the high-energy part (separated bands).

Here we want to calculate the Green's function ($\hat{G}$) of the general Hamiltonian (Eq. \ref{eq:Hamiltonian}) and we want to show that the local Green's function of the model, under certain conditions, can be expressed as a functional of the total DOS. 
Assuming that the $\vec{\sigma}$ are the Pauli matrices representing the spin part of the Hamiltonian and $\vec{\tau}$ are the orbital Pauli matrices, the Eq. \ref{eq:Hamiltonian_block} for $\boldsymbol{H}_0^{(4\times 4)}$ has the standard Weyl Hamiltonian form but with different coefficients that now are $\kvec$-dependent. Its energy dispersion function in Eq. \ref{eq:b_vector} has each component that includes the linear and the non-linear parts. We define $b_x$, $b_y$ and $M(\kvec)$ to be linear functions of $\kvec$  around $\kvec\rightarrow 0$ (nodal point). The term $M(\kvec)$ implicitly includes the linear term $b_z(\kvec$) ). With the notatiion  $\vec{b}_{\perp}(\kvec)$ and $\vec{\sigma}_{\perp}$ are $\vec{b}$ and $\vec{\sigma}$ we indicate vectors projected in the $x$,$y$ subspace. 
\\
Now, defining

\begin{align}
    &b_-(\kvec)=\left(b_x(\kvec)-ib_y(\kvec)\right)\\
    &b_+(\kvec)=\left(b_x(\kvec)+ib_y(\kvec)\right)
\end{align}
the matrix in Eq. \ref{eq:Hamiltonian_block} has the following for 

\begin{align}
    \boldsymbol{\hat{H}}_0^{(4\times 4)}(\kvec)&=\begin{pmatrix}
        M(\kvec) & b_-(\kvec) & 0 & 0 \\
        b_+(\kvec) & -M(\kvec) & 0 & 0 \\
        0 & 0 & -M(\kvec) & -b_-(\kvec) \\
        0 & 0 & -b_+(\kvec) & M(\kvec) 
    \end{pmatrix}
    \label{eq:HblockDiagonal}
\end{align}
The eigenvalues of Hamiltonian Eq. (\ref{eq:HblockDiagonal}) are easily evaluated as
\begin{equation}
    \epsilon(\kvec)=\pm \sqrt{M^2(\kvec)+b^2_x(\kvec)+b^2_y(\kvec)}.
    \label{eq:eigenvaluesHblockDiagonal}
\end{equation}

The next step is to calculate the interacting Green's function of the model. We have to notice that, if there isn't any interaction which matches the two blocks and therefore the self-energy is diagonal, we can have to deal only with a $2\times2$ matrix. The inverse Green's function reads 

\begin{align}
    \hat{G}_{\kvec}^{-1}&=\begin{pmatrix}
    z_1-M(\kvec) & -b_-(\kvec)\\
    -b_+(\kvec) & z_2+M(\kvec)\\
    \end{pmatrix}
     \label{eq:Green_Inverse}
\end{align}

where
\begin{equation}
    z_i=\omega+\mu-\Sigma_i(\omega)
    \label{eq:zeta}
\end{equation}
The self-energy is a complex quantity $\Sigma(\omega)=Re[\Sigma(\omega)]+i Im[\Sigma(\omega)]$ and each band ($i=1,2$) has in principle a different self-energy.

Finally, inverting Eq. \ref{eq:Green_Inverse}

\begin{align}
    \hat{G}_{\kvec}=&\frac{1}{ (z_2+M(\kvec))(z_1-M(\kvec))-b_\perp^2(\kvec)}\times\notag\\\notag\\
    &\times\begin{pmatrix}
    z_2+M(\kvec) & b_-(\kvec)\\
    b_+(\kvec) & z_1-M(\kvec) 
    \end{pmatrix}
    \label{eq:Green_general}
\end{align}
where $b_{\perp}(\kvec)=\sqrt{b_x^2(\kvec)+b_y^2(\kvec)}$ is the modulus of the energy dispersion in the $(x,y)$-plane.
\\\\

The single site DMFT approach involves the calculation of the local Green function  $\hat{G}_{loc}=\sum_{\kvec}\hat{G}_{\kvec}$. We will show that under some conditions the local Green function becomes a scalar quantity.

From Eq. \ref{eq:Green_general} we can deduce the equation for $\hat{G}_{loc}=\sum_{\kvec}\hat{G}_{\kvec}$ by defining the functions

\begin{align}
    &F(z_1,z_2)=\sum_{\kvec}\frac{1}{ (z_2+M(\kvec))(z_1-M(\kvec))-b_\perp^2(\kvec)}\label{eq:Ffunction}\\\notag\\
    &P(z_1,z_2)=\sum_{\kvec}\frac{M(\kvec)}{ (z_2+M(\kvec))(z_1-M(\kvec))-b_\perp^2(\kvec)}\label{eq:Dfunction}
\end{align}
and the off-diagonal terms

\begin{align}
    U_{\pm}(z_1,z_2)=\sum_{\kvec}\frac{b_{\pm}(\kvec) }{ (z_2+M(\kvec))(z_1-M(\kvec))-b_\perp^2(\kvec)}
    \label{eq:Ufunction}
\end{align}
thus we have the local $G$ as

\begin{align}
    \hat{G}_{loc}=\begin{pmatrix}
    z_2F(z_1,z_2)+P(z_1,z_2) & U_-(z_1,z_2)\\
    U_+(z_1,z_2) & z_1F(z_1,z_2)-P(z_1,z_2) 
    \end{pmatrix}
    \label{eq:Green_local_general}
\end{align}

The mass term in Eq. (\ref{eq:HblockDiagonal}) can be separated into an even and an odd term in $\kvec$ we have
\begin{equation}
    M(\kvec)=M_E(\kvec)+M_O(\kvec)
    \label{eq:splitM}
\end{equation}
where the symmetries ensure that $\sum_{\kvec} M_O(\kvec)=0$ in the first Brillouin zone. Furthermore we assume that
\begin{equation}
M_E(\kvec)=0.
\end{equation}

Let us assume also that the out-diagonal terms $b_+$ and $b_-$ are odd functions of $\kvec$ as it can be seen e.g. from tight-biding models \citep{KargarianPNAS2016}.

When $M(\kvec)$ is an odd function of $\kvec$, the function defined in Eq. (\ref{eq:Ffunction}) is even under argument exchange $F(z_1,z_2)=F(z_2,z_1)$. This property guarantees that the DMFT self-consistency admits as a possible solution $z_1=z_2=z$ and $P(z,z)=U_{\pm}(z,z)=0$. In this case the spectral function $A(\kvec,\omega)=-\frac{1}{2\pi} Im \text{Tr} \hat{G}_{\kvec}(\omega)$ from Eq. (\ref{eq:Green_general}) is  
\begin{equation}
    A(\kvec,\omega)=-\frac{1}{2\pi} Im \left(\frac{1}{z-\vert\vec{b}(\kvec)\vert}+\frac{1}{z+\vert\vec{b}(\kvec)\vert}\right)\mathbf{1}
    \label{eq:spectral}
\end{equation}
All these considerations show that the off-diagonal terms after the sum over $\kvec$ is performed vanish  and the self-consistency condition takes the following form 
\begin{align}
    \hat{G}_{loc}&=\sum_{\kvec}\frac{z}{z^2-\vert\vec{b}(\kvec)\vert^2}\mathbf{1}=\\\notag\\
    &\sum_{\kvec}\frac{1}{2}\left(\frac{1}{z-\vert\vec{b}(\kvec)\vert}+\frac{1}{z+\vert\vec{b}(\kvec)\vert}\right)\mathbf{1}
    \label{eq:Gloc_scalar}
\end{align}

The local Green's function can be written as separating the two parts: high energy ($|E(\kvec)|>\Lambda$) and low energy ($|E(\kvec)|<\Lambda$)

\begin{equation}
    \hat{G}_{loc}=\sum_{\kvec>\kvec_{\Lambda}}\hat{G}_{\kvec}^{>}+\sum_{\kvec<\kvec_{\Lambda}}\hat{G}_{\kvec}^{<}
    \label{eq:GreenLocalSeparated}
\end{equation}
where $\hat{G}_{\kvec}^{>}$ and $\hat{G}_{\kvec}^{<}$ are respectively the high and low energy limits of the Green's function in Eq. \ref{eq:Green_general} and $\vert\kvec_{\Lambda}\vert=\Lambda/v_F$.

\textit{Low energy -} The low energy part is obtained by linearization around the Weyl points 

\begin{equation}
    \hat{H}^{<}=\begin{pmatrix}
    t_zk_z & \lambda(k_x-ik_y)\\
    \lambda(k_x+ik_y) & -t_zk_z
    \end{pmatrix}
    \label{eq:H_LE_TB}
\end{equation}
where $t_z$ and $\lambda$ are the Fermi velocity near the nodal point of the $z$-component and the $x$, $y$-components respectively. The linearized Hamiltonian corresponds to the effective Weyl Hamiltonian $H^{WEYL}=v_F\kvec\cdot\vec{\tau}$  where without losing of generality we assume that the velocity is isotropic in the three directions $t_z=\lambda=v_F$.

The local Green function at low energy can be rewritten as:
\begin{equation}
    G_{loc}^{<}=\frac{1}{2}\int_{-\Lambda}^{\Lambda}d\epsilon\frac{D_{<}(\epsilon)}{z-\epsilon}
    \label{eq:Gloc_LE_DOS}
\end{equation}

where $D_{<}(\epsilon)=c\epsilon^2$ is the quadratic density of states corresponding to the linear part of the bands at low energy near the node(s) and $c$ as the normalization constant.

\textit{High Energy - } For energies larger than $\Lambda$ we consider a constant DOS model where $D^{+}_{>}$ and $D^{-}_{>}$ are respectively the electron and the hole DOSes. Since $M(\kvec)=-M(-\kvec)$ then $D^{+}_{>}=D^{-}_{>}=D_{>}=c\Lambda^2$. The term in Eq. \ref{eq:Gloc_scalar} is

\begin{align}
    &\sum_{\kvec>\kvec_{\Lambda}}\frac{1}{z\pm \vert\vec{b}(\kvec)\vert}=\\\notag\\
    =&\sum_{\kvec>\kvec_{\Lambda}}\int\frac{\delta(\epsilon\pm \vert\vec{b}(\kvec)\vert}{z-\epsilon}d\epsilon=\\\notag\\
    =&\int\frac{D_{>}^{\mp}(\epsilon)}{z-\epsilon}d\epsilon
\end{align}

The local Green's function at high energy is thus

\begin{align}
    G_{loc}^{>}&=\frac{1}{2}\left(\int_{-\infty}^{-\Lambda}d\epsilon \frac{D_{>}(\epsilon)}{z-\epsilon}+\int_{\Lambda}^{\infty}d\epsilon \frac{D_{>}(\epsilon)}{z-\epsilon}\right)
    \label{eq:Gloc_HE_DOS}
\end{align}

\textit{Final Formula - }  
We can define a total DOS from which the local Green function can be expressed as

\begin{equation}
    \hat{G}_{loc}=\int_{-\infty}^{\infty}d\epsilon\frac{\mathcal{N}(\epsilon)}{z-\epsilon}
    \label{eq:Gloc_total_DOS}
\end{equation}
where 
\begin{equation}
\mathcal{N}(\epsilon)= \begin{cases}
c\epsilon^2 \hspace{1cm}  -\Lambda<\epsilon<\Lambda\\
c\Lambda^2 \hspace{1cm} \Lambda<\vert\epsilon\vert<D
\end{cases}
\label{eq:DOSQPNSM}
\end{equation}
Notice that the normalization constant takes into account the presence of another Weyl node associated with the second block of the Hamiltonian Eq. (\ref{eq:Hamiltonian_block}) as well as spin multiplicity

\begin{equation}
\int_{-D}^{+D} \mathcal{N}(\epsilon)d\epsilon=1
\end{equation}
which gives
\begin{equation}
    c=\frac{1}{2\Lambda^2(D-\Lambda)+\frac{2}{3}\Lambda^3}
    \label{eq:c}
\end{equation}

\begin{figure}[h]
\includegraphics[width=0.5\textwidth]{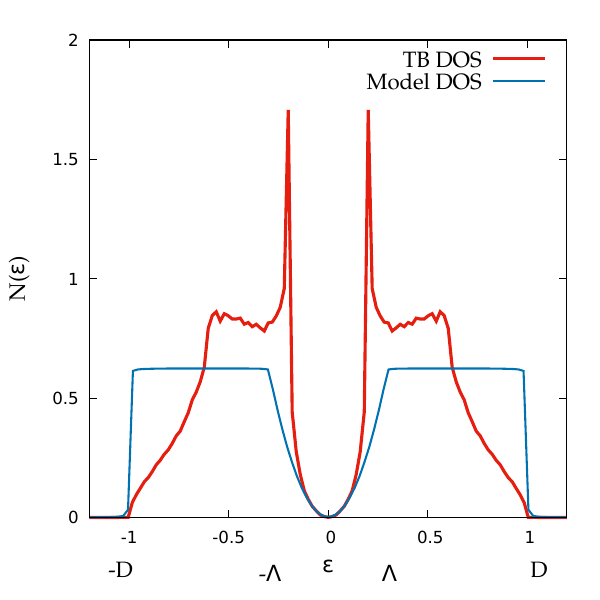}
\caption{Model of free particles DOS compared with the DOS calculated from the Tight Binding (TB) model of Ref. \citep{KargarianPNAS2016} with parameters $\lambda=0.2$. We fit our model to the curvature at the nodal point which sets our scale  $\Lambda/D=0.3$}
\label{fig:DOSdiracsquare_TB}
\end{figure}

A comparison of the present model DOS and that of the anisotropic Weyl semimetal tight binding model \citep{KargarianPNAS2016} is shown in Fig. \ref{fig:DOSdiracsquare_TB}.

\section{Kubo formula and DC conductivity}
\label{app:Kubo}
The calculation of the conductivity starts from the Kubo formula where we neglect vertex corrections
\begin{align}
    Re\sigma_{i,j}(\Omega)=&\frac{e^2\pi}{\Omega}\int_{-\infty}^{+\infty}d\omega\left[f(\omega-\mu)-f(\omega-\mu+\Omega)\right]\times\notag\\\notag\\
    &\times\int\frac{d^3k}{(2\pi)^3}\text{Tr}\left[v_i\rho(\kvec,\omega)v_j\rho'(\kvec,\omega+\Omega)\right]
    \label{eq:Kubo}
\end{align}
where $\rho(\kvec,\omega)$ is the density function defined as $\rho(\kvec,\omega)=-\frac{1}{\pi}Im[\hat{G}(\kvec,\omega)]$ and $v_i=\frac{\partial H}{\partial k_i}$ is the velocity operator. This last quantity can be calculated, for example in the $x-$direction from the upper block of Hamiltonian Eq. \ref{eq:HblockDiagonal}

\begin{equation}
    \hat{v}_x=\begin{pmatrix}
    \frac{\partial M(\kvec)}{\partial k_x} & \frac{\partial b_-(\kvec)}{\partial k_x} \\
    \frac{\partial b_+(\kvec)}{\partial k_x} & -\frac{\partial M(\kvec)}{\partial k_x}
    \end{pmatrix}
    \label{eq:Vertex_general}
\end{equation}

Instead, considering the Green's function in Eq. \ref{eq:Green_general} we can put $z_1=z_2=z$ when the components of $\vec{b}(\kvec)$ are odd respect to $\kvec$ as shown in this Appendix.
This Green's function is the same considered in \citep{Carbotte_PRB_2014} but also including the non-linear dispersion in the different components.
The $\kvec$-integral can be divided into two different sums for low and high energies

\begin{align}
      \sum_{\kvec}Tr&[v_{\kvec}\rho v_{\kvec}\rho']=\sum_{\kvec<\kvec_{\Lambda}}Tr[v_{\kvec}^{<}\rho^{<} v_{\kvec}^{<}\rho^{<'}]+\notag\\
      \notag\\
      &+\sum_{\kvec>\kvec_{\Lambda}}Tr[v_{\kvec}^{>}\rho^{>} v_{\kvec}^{>}\rho^{>'}]
       \label{eq:Trace_HE_LE_Sum}
\end{align}
where we can separate the low energy and high energy density and velocity operator.

\textit{Low Energy - }
We can calculate the vertex function in the $x-$direction from Eq. \ref{eq:Vertex_general} assuming the limit $k_x\rightarrow 0$ near the Weyl point of Eq. \ref{eq:H_LE_TB} and considering an isotropic Fermi velocity $\lambda=t_z=v_F$,

\begin{equation}
    \hat{v}_x^{<}=\begin{pmatrix}
    0 & v_F\\
    v_F & 0
    \end{pmatrix}=v_F\vec{\sigma_x}.
    \label{eq:VertexLE}
\end{equation}
It has the same spinorial structure of that e.g. Ref. \citep{Carbotte_PRB_2014}
including intraband and interband transitions.

\textit{High Energy - } Again, the high energy contribution to the conductivity can be calculated by neglecting the interband-contribution terms  $b_{\pm}$ both in the vertex function (Eq. \ref{eq:Vertex_general}) and in the Green's function (Eq. \ref{eq:Green_general}  with $z_1=z_2$). This means that the vertex function in the high-energy regime,  $\hat{v}_x^{>}$, looks
\begin{equation}
    \hat{v}_x^{>}=\begin{pmatrix}
    \frac{\partial M(\kvec)}{\partial k_x} & 0\\
    0 & -\frac{\partial M(\kvec)}{\partial k_x}
    \end{pmatrix}=\frac{\partial M(\kvec)}{\partial k_x}\vec{\sigma_z}
    \label{eq:VertexHE}
\end{equation}
Both the velocity operator and the Green's function are diagonal therefore the two orbitals are independent.
The spectral density $\rho^>$ is

\begin{align}
    \rho^>(\kvec,\omega)=\begin{pmatrix}
    \rho_+ & 0 \\
    0 & \rho_-
    \end{pmatrix}(\kvec,\omega)
    \label{eq:rhoHE}
\end{align}
where

\begin{align}
    \rho_{\pm}(\kvec,\omega)=-\frac{1}{\pi} \frac{Im[\Sigma]}{(\left[\omega{\mp}M(\kvec)-Re[\Sigma]\right]^2+\left[Im[\Sigma]\right]^2)}
\end{align}

The Trace become

\begin{equation}
    \text{Tr}\left[\hat{v}^{>}_x\hat{\rho}^{>}\hat{v}^{>}_x\hat{\rho}^{>'}\right]=\left(\frac{\partial M(\kvec)}{\partial k_x}\right)^2(\rho_+\rho_+'+\rho_-\rho_-')
\end{equation}
and contains only the intraband transition terms.

Therefore, the DC conductivity is given by $\lim_{\Omega\rightarrow 0}Re\sigma(\Omega)$
\begin{align}
    Re\sigma_{xx}&=e^2\pi\int_{-\infty}^{+\infty}d\omega\left(-\frac{df}{d\omega}\right)\times\notag\\
    &\times\int_{\kvec>\kvec_{\Lambda}}\frac{d^3k}{(2\pi)^3}\phi(\kvec)\left[\rho_+\rho_++\rho_-\rho_-\right]
\end{align}
where $\phi(\kvec)$ is the vertex function and $\kvec_{\Lambda}=\Lambda/v_F$ as in the Appendix \ref{app:DMFT}.
Since we can define a vertex density as $\phi(\epsilon)D(\epsilon)=\sum_{\kvec}\phi(\kvec)\delta(\epsilon-\epsilon_{\kvec})$, then $\int\frac{d^3k}{(2\pi)^3}\phi(\kvec)=\int d\epsilon \phi(\epsilon)D(\epsilon)$ and the integral in $\kvec$ is

\begin{align}
     &Re\sigma_{xx}=e^2\pi\int_{-\infty}^{+\infty}d\omega\left(-\frac{df}{d\omega}\right)\times\notag\\
     &\left[\int_{\Lambda}^{+D}\phi(\epsilon)D_{>}^+(\epsilon)\rho^{(+)}\rho^{(+)}+\int_{-D}^{-\Lambda}\phi(\epsilon)D_{>}^-(\epsilon)\rho^{(-)}\rho^{(-)}\right]  
     \label{eq:Sigma_highEnergy}
\end{align}

where this time $\rho^{(\pm)}=\rho^{(\pm)}(\epsilon,\omega)$ are 

\begin{align}
    \rho^{(\pm)}(\epsilon,\omega)=-\frac{1}{\pi} \frac{Im[\Sigma]}{(\left[\omega{\mp}\epsilon-Re[\Sigma]\right]^2+\left[Im[\Sigma]\right]^2)}
\end{align}

This formulation is valid whenever $M^2(\kvec) \gg (b^2_x+b^2_y)$.

Now, we need to briefly discuss the evaluation of $\phi(\epsilon)$. Our objective is to match the high-energy vertex function, $\phi(\epsilon)$, with the constant vertex described in Ref. \citep{Carbotte_PRB_2014}, which is $\phi_{LE}=v_F^2/3$. To simplify the analysis, we assume that the vertex function in the high-energy case is constant with respect to energy and we force both to be the same constant $\phi(\epsilon)=\phi_{LE}=1/3$.

\textit{Final Formula - } Taking into account the parts in Ref. \citep{Carbotte_PRB_2014} with the appropriate considerations for the vertex function and Eq. \ref{eq:Sigma_highEnergy} and following Eq. \ref{eq:Trace_HE_LE_Sum} 
\begin{align}
    Re\sigma_{xx}&=e^2\pi\int_{-\infty}^{+\infty}d\omega\left(-\frac{df}{d\omega}\right)\times \notag\\\notag\\
   &\times(\int_{-D}^{D}d\epsilon\mathcal{N}(\epsilon)\left[\rho^{(+)}\rho^{(+)}+\rho^{(-)}\rho^{(-)}\right]+ \notag\\\notag\\  
    &+\int_{-\Lambda}^{\Lambda}d\epsilon \hspace{2pt} 2 c\epsilon^2 \left[\rho^{(+)}\rho^{(-)}+\rho^{(-)}\rho^{(+)}\right])
    \label{eq:condtotal}
\end{align}
Here, $\mathcal{N}(\epsilon)$ and $c$ are the same as in Eq. (\ref{eq:DOSQPNSM}).

\section{Quasiparticle Damping in weak coupling}
\label{app:damping}

\begin{figure}[t]
\includegraphics[scale=0.8]{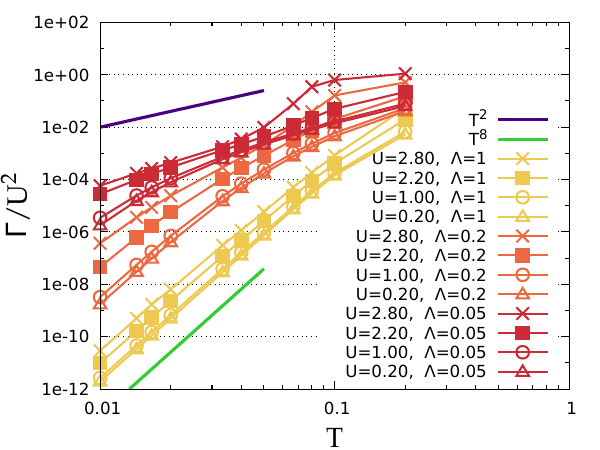}
\centering
\caption{The quasi-particle damping at half-filling normalized with $U^2$ as a function of the temperature ($T=T/D$) for $\Lambda=\Lambda/D=1.0$, $ 0.2$, $0.05$ and several values of $U=U/D$ from the metallic side approaching the MIT.}
\label{fig:GammaLambdavsT}
\end{figure}

\begin{figure}[t]
\includegraphics[scale=0.8]{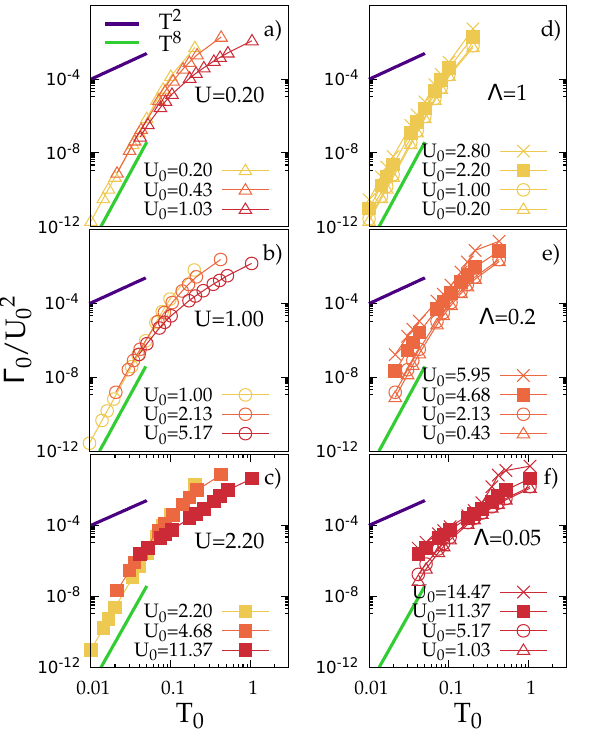}
\centering
\caption{The quasi-particle damping at half-filling normalized with $U_0^2$ at fixed $U=U/D$ and $\Lambda=\Lambda/D$ as a function of the temperature ($T/D_0$) and scaled to obtain a comparison between quantities with the same bare Fermi velocity. The subscript $0$ indicates that the quantity in units $D_0$ [see Appendix \ref{app:scaling}]. The colour code is the same as in Fig. \ref{fig:GammaLambdavsT}}
\label{fig:GammaLambdavsTScaled}
\end{figure}

From perturbation theory calculations  \citep{wagner}
the small energy and temperature behaviour of the imaginary part of the self-energy has a polynomial dependence
\begin{equation}
Im\left[\Sigma(\omega)\right]=-8\pi^9 T^8 U^2\left(\frac{1}{2\pi^2v_F^3}\right)^3 P_8(\omega/ T)
\label{eq:NSMImaginarySelf}
\end{equation}
with
\begin{align}
& P_8(x)=a_0+a_2x^2+a_4 x^4+a_8 x^8\\
& a_0= \frac{3\pi^8}{128} \\
& a_2= \frac{31\pi^6}{504\cdot2!}\\
& a_4=\frac{7\pi^4}{40\cdot4!} \\
& a_8=\frac{1}{8!}
\end{align}
and according to our definition of the density of states in Eqs. \ref{eq:DOSQPNSM}, \ref{eq:c}, $v_F^3=1/(2\pi^2c)=\frac{1}{2\pi^2}(2\Lambda^2(D-\Lambda)+\frac{2}{3}\Lambda^3)$.

In Fig. \ref{fig:GammaLambdavsT} is shown the temperature evolution  of the quasi-particle damping $\Gamma = -\text{Im}[\Sigma(\omega=0)]$ normalized with $U^2$, for different values of the interaction strength ($U=U/D$) and for different values of the cutoff ($\Lambda/D$). Temperature exponent ranges from a pure $T^8$ characteristic of nodal semimetal to $T^2$ being the characteristic exponent of the Fermi liquid phase. 
As far as $\Lambda/D<1$, non-linear band effects extend to low energy and the exponent decreases from 8 to 2. Increasing $U/D$, for $\Lambda/D<1$, leads to a more evident decrease of the exponent from the second-order perturbation theory result. A clear deviation from perturbation theory is shown for the smallest value of $\Lambda/D$
and the largest value of $U$.

Employing a suitable scaling of the $\Gamma$ and $T$ in order to compare systems with the same zero energy Fermi velocity (see Appendix \ref{app:scaling}) we get the rescaled curves shown in  Fig. \ref{fig:GammaLambdavsTScaled}. 
We identify a clear  crossover from the semimetal liquid to the Fermi liquid behaviour of the quasi-particle damping which occurs when the $T\sim 0.1 v_F$. For the smallest $\Lambda/D=0.05$ the range of $T$ is too narrow to observe the semimetal liquid perturbative behaviour which we expect at very low temperatures.

\section{Scaling at weak coupling}
\label{app:scaling}
\begin{figure}[h]
\begin{tikzpicture}
    \node[inner sep=0pt] (image1) at (0,0) {\includegraphics[width=0.45\textwidth]{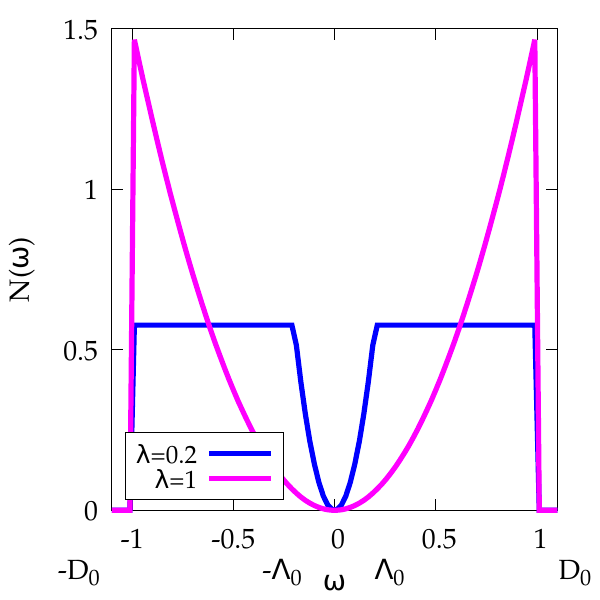}};
    \node[inner sep=0pt] (image2) at (0.45,2.8) {\includegraphics[scale=0.4]{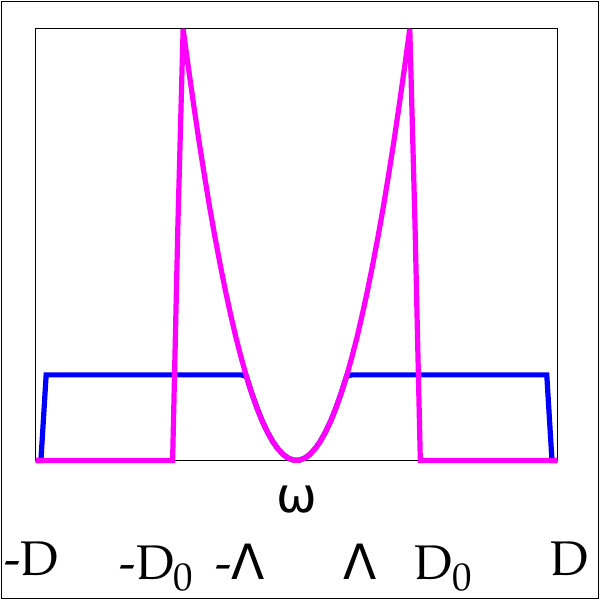}};
\end{tikzpicture}
\centering
\caption{Free particle DOS used in the IPT codes. Notice different curvature around $\omega=0$ which imply different bare Fermi velocity. In the inset: Scaled free particle DOS which shows the same bare Fermi velocity.
}
\label{fig:DOSScaling}
\end{figure}

In Fig. \ref{fig:DOSScaling} we show the total non-interacting DOS for $\Lambda/D=1$ and for $\Lambda/D<1$.  In order to compare systems with the same {\it bare} Fermi velocity (curvature of the DOS around zero energy) a rescaling of the DOS (inset of Fig. \ref{fig:DOSScaling}) is required. The rescaling affects both temperatures and the interaction strengths for $\Lambda/D<1$.

Firstly, we fix the half-bandwidth of the model with $\Lambda/D=1$ to a value $D_0$. The normalisation constant entering the DOS in Eq. \ref{eq:DOSQPNSM} as $\Lambda=D_0$ can be expressed as 
\begin{equation}
c=\frac{3}{2D_0^3}
\end{equation}
Then, we can define a scaling factor such that
\begin{equation}
c_0S^3(\lambda)=\frac{3}{2D_0^3}    
\end{equation}
where the subscript $0$ stands for the constants describing the DOS model with half-bandwidth $D_0$. The constant $\lambda=\Lambda/D=\Lambda_0/D_0$ is independent from the rescaling.
Recalling the formula for the normalization constant in \ref{eq:c},  $c_0=1/\left[(\frac{2}{3}\lambda^3+2\lambda^2(1-\lambda))D_0^3\right]$ and therefore
\begin{equation}
    S(\lambda)=(\lambda^3+3\lambda^2(1-\lambda))^{1/3}.
\end{equation}
In order to keep $D_0$ constant we should rescale the half-bandwidth as
\begin{equation}
    D=\frac{D_0}{S(\lambda)}
\end{equation}
but also the temperature and the interaction consistently
\begin{align}
& \frac{T}{D_0}=\frac{T}{D}\frac{D}{D_0}=\frac{T}{S(\lambda)}>\frac{T}{D} \\ \notag \\
&\frac{U}{D_0}=\frac{U}{D}\frac{D}{D_0}=\frac{U}{S(\lambda)}>\frac{U}{D}.
\end{align}

\section{Optical Conductivity}
\label{app:optcond}

\begin{figure}[h]
\begin{tikzpicture}
    \node[inner sep=0pt] (image1) at (0,0) {\includegraphics[width=0.45\textwidth]{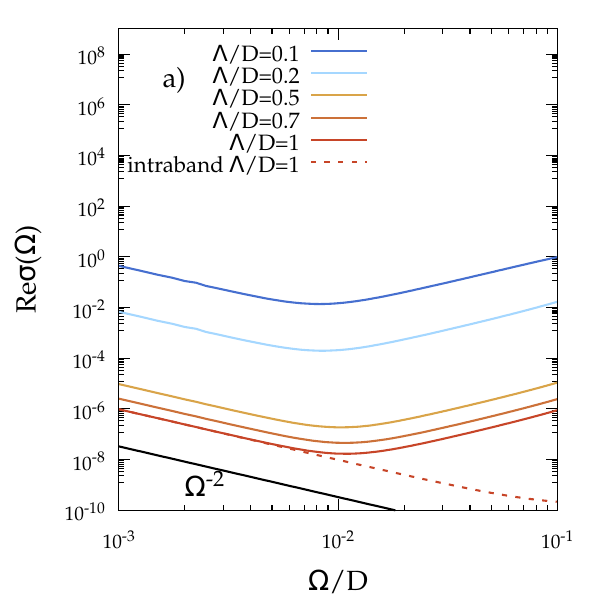}};
    \node[inner sep=0pt] (image2) at (2.3,2.2) {\includegraphics[scale=0.33]{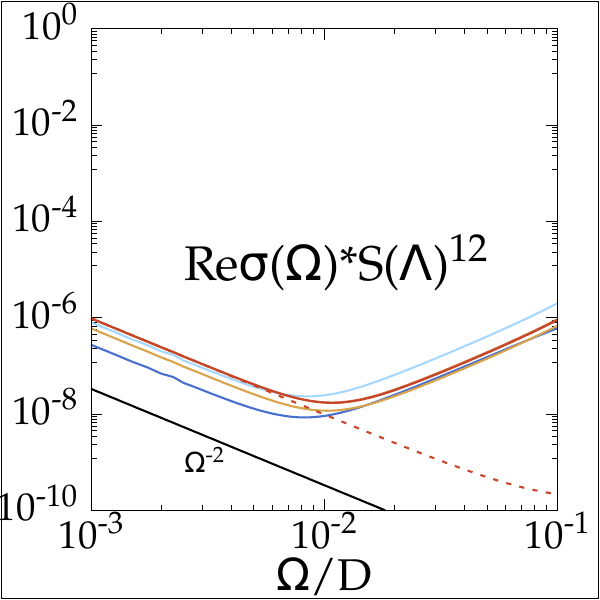}};
\end{tikzpicture}
\quad\includegraphics[width=0.49\textwidth]{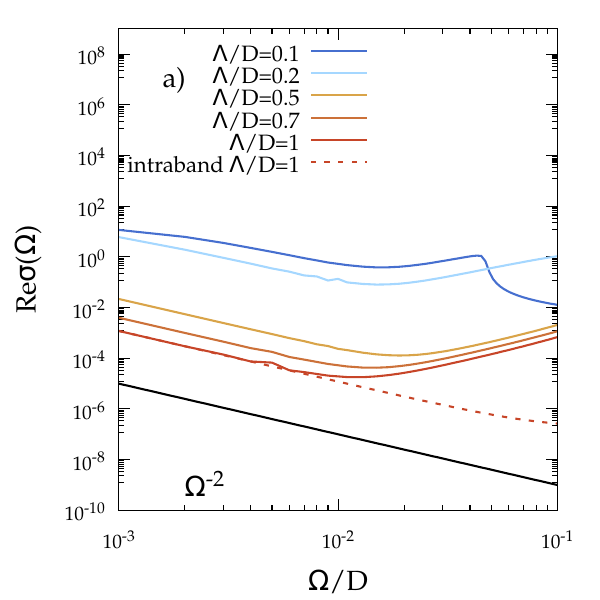}
\centering
\caption{Optical conductivity at half-filling for $T/D=0.02$ and $U/D=0.5$ (a) and $U/D=3.0$ (b). In both panels, the dashed red curve indicates the intraband contribution to the optical spectra at $\Lambda/D=1$. They are obtained by taking into account only the intraband contribution from the Kubo formula (Eq. \ref{eq:Kubo}).
}
\label{fig:Opt_cond}
\end{figure}

Using the Kubo formula \ref{eq:Kubo} the optical conductivity is

\begin{align}
    Re\sigma_{xx}&=e^2\pi\int_{-\infty}^{+\infty}d\omega\left(-\frac{df}{d\omega}\right)\times \notag\\\notag\\
   &\times(\int_{-D}^{D}d\epsilon\mathcal{N}(\epsilon)\left[\rho^{(+)}\rho'^{(+)}+\rho^{(-)}\rho'^{(-)}\right]+ \notag\\\notag\\  
    &+\int_{-\Lambda}^{\Lambda}d\epsilon \hspace{2pt} 2 c\epsilon^2 \left[\rho^{(+)}\rho'^{(-)}+\rho^{(-)}\rho'^{(+)}\right])
    \label{eq:optcondtotal}
\end{align}
where $\rho'$ is the spectral function evaluated at $\omega+\Omega$ and $\rho$ at $\omega$.
The results from the Kubo formula (Eq. \ref{eq:optcondtotal}) are shown in Fig. \ref{fig:Opt_cond} in the weak (a) and strong coupling (b) NSM phases. Note that, the intraband contribution is obtained from Eq. \ref{eq:optcondtotal} taking into account only the first term in the integral over $\epsilon$.

Note that, our approximation for the optical conductivity \ref{eq:optcondtotal} neglects the interband transitions at large energy $\Omega$ and therefore can describe accurately the optical spectra at low energy  $\Omega<\Lambda$. A sharp zero frequency Lorentzian Drude peak is indeed found in Fig. \ref{fig:Opt_cond}, and its $\Omega^{-2}$ tail is shown. Only intraband transitions contribute to this peak. Both the amplitude and the width of this peak strongly depend on the energy scale $\Lambda$. At an energy $\omega > T$ as $T<\Lambda$ interband transitions start to contribute producing an increase in optical absorption. 
Since at weak coupling the intraband absorption is of the Drude form
\begin{equation}
    \sigma_{intra}(\Omega)=\frac{\sigma(0) \Gamma^2_{opt}}{\Omega^2+\Gamma^2_{opt}}    
\end{equation}
the effect of $\Lambda/D <1$ reduces at low temperature to simple scaling of low-frequency optical conductivity since for $\Omega>\Gamma_{opt}$ the scaling of the optical spectra reduces to the scaling of $\sigma(0) \Gamma^2_{opt}$. From the results shown in Fig. \ref{fig:Rho_vs_T} (d) we know that $\sigma(0)$ scales with $S^6(\Lambda)$. On the other hand, as it is shown in Fig. \ref{fig:OptGamma} and derived below  $\Gamma_{opt}$ scales as the quasiparticle damping ($S^{-9}(\Lambda)$) and therefore at weak coupling and low temperature the scaling of low energy optical conductivity is that of $S^{-12}(\Lambda)$. The scaled optical conductivity is shown in the inset of Fig. \ref{fig:Opt_cond} (a). In the strongly correlated NSM phase, the scaling does not hold but still, the low-frequency optical conductivity has a Lorentzian form with $\Gamma_{opt}$ that deviates from a simple proportionality with quasiparticle damping as shown in Fig. \ref{fig:OptGamma} of the main text.

We now derive the relation between $\Gamma_{opt}$ and $\Gamma$. By using the intraband part of the Kubo formula in Eq. \ref{eq:optcondtotal}, each term of the optical conductivity can be written as the integral over energies and frequencies of the product of two Lorentzians

\begin{align}
     &Re\sigma_{i,j}(\Omega)=\frac{e^2\pi}{\Omega}\int_{-\infty}^{+\infty}d\omega\left[f(\omega-\mu)-f(\omega-\mu+\Omega)\right]\times \notag \\ \notag \\
    &\times\int d\epsilon \mathcal{N}(\epsilon) \frac{\Gamma(\omega)}{(\omega-Re\Sigma(\omega)-\epsilon)^2+\Gamma^2(\omega)}\cdot \notag \\ \notag \\
    &\cdot \frac{\Gamma(\omega+\Omega)}{(\omega+\Omega-Re\Sigma(\omega+\Omega)-\epsilon)^2+\Gamma^2(\omega+\Omega)} \\ \notag \\
    &=\frac{e^2\pi}{\Omega}\int_{-\infty}^{+\infty}d\omega\left[f(\omega-\mu)-f(\omega-\mu+\Omega)\right]\times I(\omega,\Omega)
    \label{eq:OPT_COND_Kubo_energy}
\end{align}

Let $\Gamma(\omega)=\Gamma_1$, $\Gamma(\omega+\Omega)=\Gamma_2$ and  $\omega-Re\Sigma(\omega)=\epsilon_1$, $\omega+\Omega-Re\Sigma(\omega+\Omega)=\epsilon_2$, then the integral in energy $\epsilon$ can be made using the specific model with $\Lambda/D=1$ and the corresponding quadratic density of states $\mathcal{N}(\epsilon)=3\epsilon^2/2D^3$

\begin{align}
    I(\omega,\Omega)& =\int d\epsilon \mathcal{N}(\epsilon) \frac{\Gamma_1}{(\epsilon_1-\epsilon)^2+\Gamma^2_1} \cdot \frac{\Gamma_2}{(\epsilon_2-\epsilon)^2+\Gamma^2_2} \notag \\ \notag \\
    & =\frac{3\pi}{2D^3}\frac{\Gamma_1\epsilon_2^2+\Gamma_2\epsilon_1^2+\Gamma_1^2\Gamma_2+\Gamma_2^2\Gamma_1}{(\epsilon_1-\epsilon_2)^2+(\Gamma_1+\Gamma_2)^2} 
    \label{eq:OPT_COND_Integral_energy}
\end{align}

Hence, we perform some approximations: i) we neglect the real part of the self-energy  ii) we take a constant imaginary part of the self-energy in the limit of $\omega\ll T$ and $\Omega\ll T$  ($\Gamma_1=\Gamma_2=\Gamma$).

The optical conductivity is thus

\begin{align}
    Re\sigma_{i,j}(\Omega)&=\frac{3\pi^2}{2D^3}\int_{-\infty}^{+\infty}d\omega\left(-\frac{\partial f}{\partial \omega}\right)\cdot \notag \\ \notag \\ 
    & \cdot \frac{\Gamma(\omega^2+(\omega+\Omega)^2)+2\Gamma^3}{\omega^2+(2\Gamma)^2} 
\end{align}

since the derivative of the Fermi function can be written as

\begin{equation}
    -\frac{\partial f(\omega)}{\partial \omega} = \frac{1}{2}\frac{\beta}{1+cosh(\beta\omega)}
\end{equation}

then the integral over the frequencies is

\begin{equation}
    Re\sigma(\Omega)=\frac{3\pi}{2D^3}\frac{\Gamma}{\Omega^2+(2\Gamma)^2}\left(\frac{2\pi}{3\beta^2}+2\Gamma^2+\Omega^2\right)
\end{equation}

In the range $\Gamma\ll \Omega,\omega\ll T$

\begin{equation}
    Re\sigma(\Omega)=\frac{\pi^3}{D^3\beta^2}\frac{\Gamma}{\Omega^2+(2\Gamma)^2}
\end{equation}

then the optical damping is twice as the quasiparticle damping ($\Gamma_{opt}=2\Gamma$).
\\

\section{DOS at non-zero chemical potential}
\label{app:Filling}

\begin{figure}[h]
\includegraphics[width=0.49\textwidth]{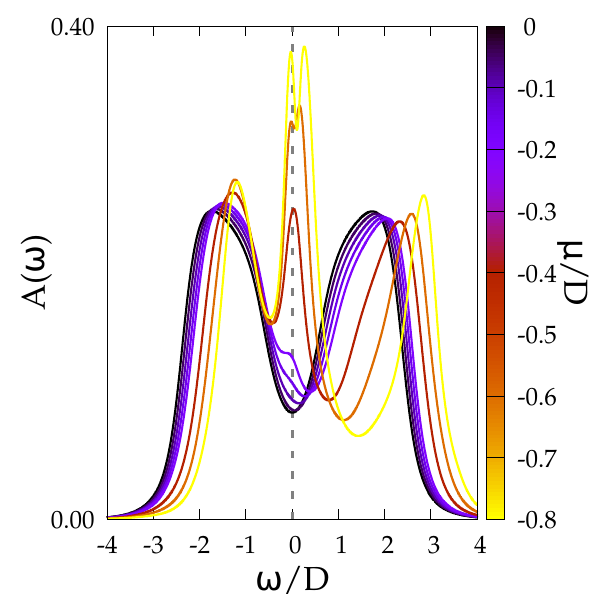}
\centering
\caption{The spectral function shown at various chemical potentials for $\Lambda/D=0.2$, $U/D=3.0$ and  $T=T^*(\mu=0.0)=0.1D$ where at half-filling the semimetal behaviour (U-shape) disappears. The dashed grey line indicates the position of the Fermi level.
}
\label{fig:nohalffilling}
\end{figure}

For finite doping, we have used a modified version of the IPT proposed in \citep{kajueter_new_1996} using the formulation illustrated in \citep{wagner}.
As it happens within DMFT in the metallic Hubbard model the temperature $T^*$ at which the central U-shape gets destroyed depends strongly on chemical potential \citep{georges_dynamical_1996}.
In Fig. \ref{fig:nohalffilling} is shown the evolution of the DOS by varying the chemical potential at constant temperature $T^*/D(\mu=0.0)=0.1$. Moving away from the nodal point a rebirth of the central U-shape is observed.

\section{Second-order perturbation theory calculations}
\label{app:secondordercalc}

The imaginary part of the self-energy for a generic DOS, calculated with the second-order perturbation theory, is
\begin{equation}
\begin{split}
Im\left[\Sigma(\omega)\right]&=-\pi U^2 \int d\epsilon_{k}d\epsilon_{k'}d\epsilon_{k''}\mathcal{N}(\epsilon_{\vec{k}})\mathcal{N}(\epsilon_{\vec{k}'})\mathcal{N}(\epsilon_{\vec{k}''})\cdot\\
&\cdot\left[f(\epsilon_{\vec{k}'})(1-f(\epsilon_{\vec{k}''}))(1-f(\epsilon_{\vec{k}}))\right. +\\
&\left. +(1-f(\epsilon_{\vec{k}'}))f(\epsilon_{\vec{k}''})f(\epsilon_{\vec{k}})\right] \cdot\\
&\cdot\delta(\omega+\epsilon_{\vec{k}}-\epsilon_{\vec{k}'}+\epsilon_{\vec{k}''})
\end{split}
\end{equation}

as calculated in \citep{wagner}.

In \citep{wagner} was calculated the specific imaginary part of the self-energy for a quadratic dispersed DOS. Here, the calculation for the flat dispersed DOS is shown as it is relevant for our  $\Lambda/D=0$ case. 

We consider a flat band with large bandwidth $D$ which gives 
\begin{equation}
Im\left[\Sigma(\omega)\right]=-\frac{\pi U^2}{(2D)^3}\left(\xi(\omega)+\xi(-\omega)\right)
\end{equation}
with
\begin{align}
\xi(\omega)&=\int d\epsilon_{k}d\epsilon_{k'}d\epsilon_{k''}\delta(\omega-(\epsilon_{k}+\epsilon_{k'}+\epsilon_{k''}))f(\epsilon_{\vec{k}})
f(\epsilon_{\vec{k'}})f(\epsilon_{\vec{k''}})\notag \\
&=\int \frac{d\alpha}{2\pi}e^{i\alpha\omega}A^3(\alpha)
\end{align}
$A(\alpha)$ is given by
\begin{equation}
A(\alpha)=\int d\epsilon \hspace{0.1cm}e^{-i\tilde{\alpha}\epsilon}f(\epsilon)
\end{equation}
where $\tilde{\alpha}=\alpha+i\delta$. A complex integration is then performed.
\\

When $\alpha\geq0$ we close the integration path in the lower complex half-plane
\begin{align}
A(\alpha)&=i2\pi T\hspace{0.1cm}\sum_{m=0}^{\infty}e^{i\tilde{\alpha}(-i\pi T (2m+1))}\notag \\
&=i\pi T \frac{1}{\hspace{0.1cm}sinh(\tilde{\alpha}\pi T)}
\end{align}
if $\alpha<0$ we obtain the same result closing the path in the upper half plane
\begin{equation}
A^3(\alpha)=-i(\pi T)^3\frac{1}{sinh^3(\tilde{\alpha}\pi T)}.
\end{equation}
We expand around the poles $\tilde{\alpha}=\frac{in}{T}$ with $n\in\mathbb{N}$
\begin{align}
A_n^3(\alpha)&=-i(\pi T)^3\left[\frac{1}{(\pi T)^3(\tilde{\alpha}-\frac{in}{T})^3}-\frac{1}{2(\pi T)(\tilde{\alpha}-\frac{in}{T})}\right]\notag\\
&=-i(-1)^n\left[\frac{1}{(\tilde{\alpha}-\frac{in}{T})^3}-\frac{(\pi T)^2}{2(\tilde{\alpha}-\frac{in}{T})}\right]
\end{align}
which gives
\begin{align}
\xi(\omega)&=\int \frac{d\alpha}{2\pi}e^{i\alpha\omega}A^3(\alpha)=\oint\frac{dz}{2\pi}e^{iz\omega}A^3(z)=\notag \\ \notag \\
&=i(-i)\sum_{n=1}^{\infty}(-1)^ne^{-\frac{n\omega}{T}}\left[-\frac{\omega^2}{2!}-\frac{(\pi T)^2}{2}\right]=\notag \\ \notag \\
&=\frac{1}{1+e^{\frac{\omega}{T}}}\left[\frac{\omega^2}{2}+\frac{(\pi T)^2}{2}\right]
\end{align}
We get for the imaginary part of the self-energy
\begin{align}
Im\left[\Sigma(\omega)\right]&=-\frac{\pi U^2}{(2D)^3}\left(\xi(\omega)+\xi(-\omega)\right)=\notag \\
&=-\frac{\pi U^2}{(2D)^3}\frac{1}{2}\left[\omega^2+(\pi T)^2\right]
\label{eq:ImmSigmaSquare}
\end{align}

The real part of the self-energy is obtained using the Kramers-Kroening relations using a large frequency cutoff $L$ as
\begin{equation}
Re\left[\Sigma(\omega)\right]=-\frac{1}{\pi}P\int_{-L}^{L}d\epsilon\frac{Im\left[\Sigma(\epsilon)\right]}{\omega-\epsilon}.
\end{equation}
We use the  Eq. \ref{eq:ImmSigmaSquare} 
\begin{equation}
-Im\left[\Sigma(\omega)\right]=\left[a_2\omega^2+a_0T^2\right]
\end{equation}
where 
\begin{equation}
\begin{split}
&a_0=\frac{U^2\pi^3}{2(2D)^3}\\
&a_2=\frac{U^2\pi}{2(2D)^3}
\end{split}
\end{equation}

For $\omega<<L$ we get
\begin{equation}
\frac{Re\left[\Sigma(\omega)\right]}{D}\simeq-\frac{U^2}{D^2}\left[\frac{\pi^2 L}{8D^2}-\frac{T^2}{8D^2 L}\right]\omega
\end{equation}
which shows a linear dependence in $\omega$.

\section{Chemical potential dependence of the filling}
\label{app:nvsmu}

\begin{figure}[h]
\includegraphics[width=0.45\textwidth]{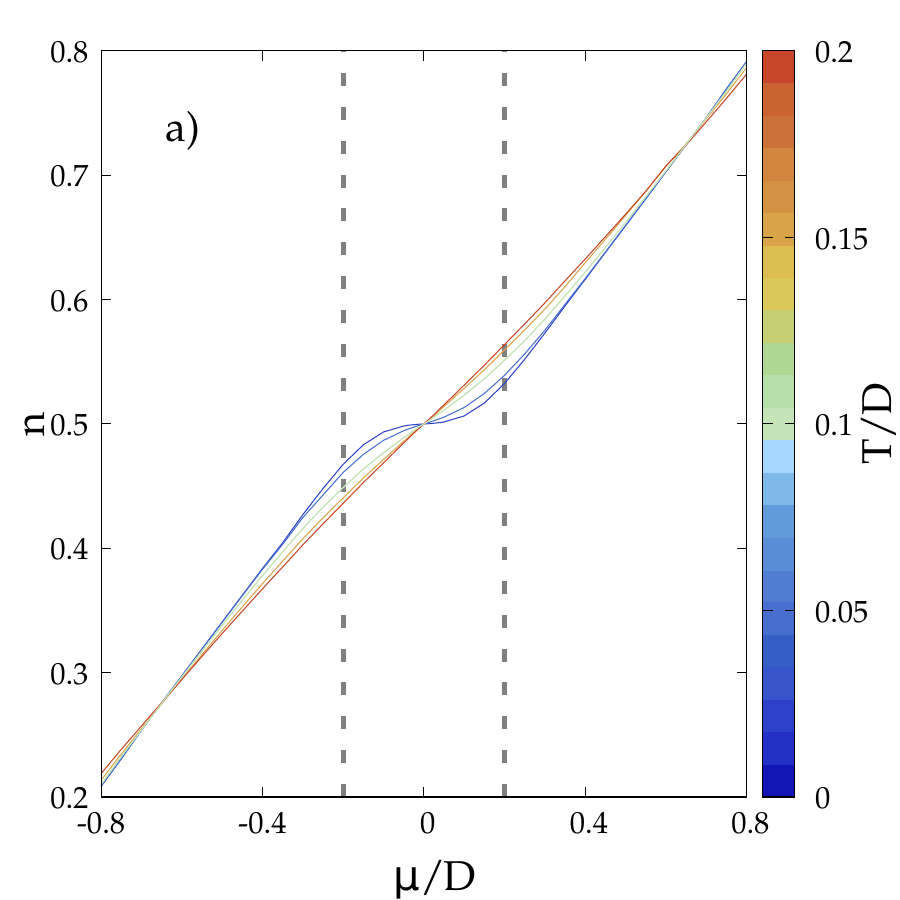}
\quad\includegraphics[width=0.45
\textwidth]{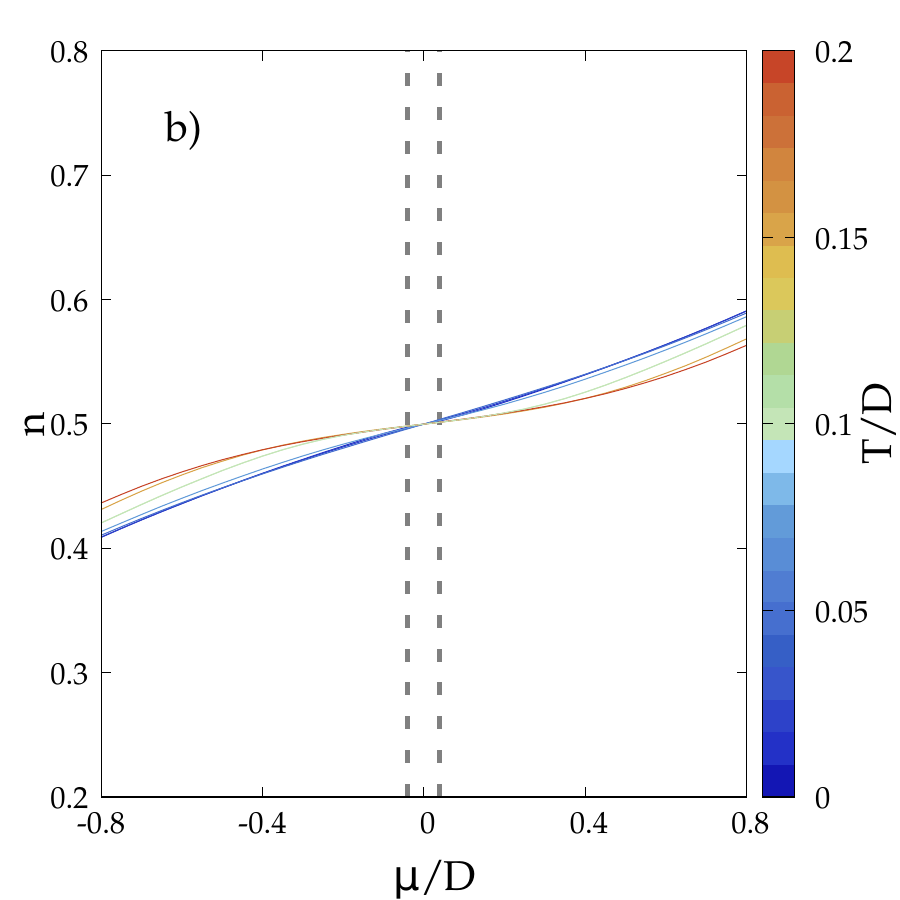}
\centering
\caption{Filling as a function of the chemical potential for different temperatures at $\Lambda/D=0.2$ and $U/D=0.5$ (a) and $U/D=3.0$ (b). Dashed grey lines indicate the width of the U-shaped structure in the spectral function.
}
\label{fig:nvsmu}
\end{figure}

The compressibility shown in figures \ref{fig:Nernstmu_WeakCoupling} and \ref{fig:Nernstmu_StrongCoupling} are obtained by numerical derivative of the filling with respect to the chemical potential. In Fig. \ref{fig:nvsmu} a) and b) the dependence of the filling on the chemical potential is shown explicitly. The small density of carriers in the node region is at the origin of the drop in compressibility near the node.

\end{document}